\documentclass[aps,prl,reprint,groupedaddress,showpacs,amsmath,amssymb,twocolumn,floatfix]{revtex4-1}
\usepackage{graphicx}
\usepackage{bm}
\usepackage{color}
\usepackage[normalem]{ulem}
\usepackage{cleveref}
\usepackage{amsmath}

\newcommand{\Bpara}{$B_{\parallel}$}

\newcommand{\muu}{$\mu_{tTLG}$}

\newcommand{\dmudnu}{$d\mu/d\nu$}

\newcommand{\IF}{IF$_3$}

\begin{document}

\title{Coulomb screening and thermodynamic measurements in magic-angle twisted trilayer graphene}


\author{Xiaoxue Liu$^{1}$}
\author{Naiyuan James Zhang$^{1}$}
\author{K. Watanabe$^{2}$}
\author{T. Taniguchi$^{2}$}
\author{J.I.A. Li$^{1}$}
\email{jia\_li@brown.edu}

\affiliation{$^{1}$Department of Physics, Brown University, Providence, RI 02912, USA}
\affiliation{$^{2}$National Institute for Materials Science, 1-1 Namiki, Tsukuba 305-0044, Japan}

\date{\today}

\maketitle

\textbf{The discovery of magic-angle twisted trilayer graphene (tTLG) adds a new twist  to the family of graphene moir\'e. The additional graphene layer unlocks a series of intriguing properties in the superconducting phase, such as the violation of Pauli limit and re-entrant superconductivity at large in-plane magnetic field. In this work, we integrate magic-angle tTLG into a double-layer structure to study the superconducting phase. Utilizing proximity screening from the adjacent metallic layer, we examine the stability of the superconducting phase and demonstrate that Coulomb repulsion competes against the mechanism underlying Cooper pairing. Furthermore, we use a combination of transport and thermodynamic measurements to probe the isospin order, which points towards a spin-polarized and valley-unpolarized isospin configuration at half moir\'e filling, and for the nearby fermi surface. Our findings provide important constraints for theoretical models aiming to understand the nature of superconductivity. A possible scenario is that electron-phonon coupling stabilizes a superconducting phase with a spin-triplet, valley singlet order parameter. }

Graphene moir\'e structures provide a paradigm system to study correlated physics and superconductivity in the 2D limit. A simple twist between graphene layers is shown unlock a rich phase space, where correlation driven insulators ~\cite{Liu2020DBLG,Cao2020DBLG,Cao2018a} coexist with superconductivity ~\cite{Cao2018b,Yankowitz2019SC,Lu2019SC} and ferromagnetism ~\cite{Sharpe2019,Serlin2019,Polshyn20201N2,Chen2020ABC,Chen20201N2}. 
As an important step towards a better microscopic understanding of the superconducting phase in graphene moir\'e, it is recently demonstrated that Cooper pairing in magic-angle twisted bilayer graphene (tBLG) competes against the influence of Coulomb repulsion between charge carriers ~\cite{Liu2021DtBLG,Stepanov2019interplay,Saito2019decoupling}, suggesting that superconductivity in the bilayer moir\'e likely arises from electron-phonon coupling. At the same time, a Pomeranchuk-type phase transition is observed in tBLG, where the high temperature phase is associated with the large electronic entropy and fluctuating isospin moments ~\cite{Saito2021pomeranchuk, Rozen2021pomeranchuk}. The Pomeranchuk effect points towards small isospin stiffness in the moir\'e band of tBLG. These observations set the stage for the intriguing open questions regarding the nature of superconductivity in graphene moir\'e systems. 


Superconductivity is recently reported in twisted trilayer graphene at the predicted magic angle, adding a new and intriguing member to graphene moir\'e structures. This three-layered structure is shown to be highly versatile, as band dispersion and Fermi surface contour are both suggested to be tunable with a perpendicular electric field $D$ ~\cite{Park2021tTLG,Hao2021tTLG}. 
Interestingly, the superconducting phase in magic-angle tTLG remains robust against a large in-plane magnetic field, which violates the Pauli limit  for conventional spin-singlet superconductors ~\cite{Cao2021Pauli}. This observation provides experimental support for spin-triplet pairing, where two electrons in a Cooper pair share the same spin quantum number. Such pairing symmetry has been previously observed in a range of unconventional superconductors, such as the Anderson-Brinkman-Morel phase of Helium-3 superfluid ~\cite{Vollhardt1990}, Ru$_2$SO$_4$ ~\cite{Mackenzie2003} and UPt$_3$ ~\cite{Schemm2014UPt,Strand2010UPt}. However, graphene moir\'e structures feature a spin-valley $SU_4$ symmetry, requiring us to consider both isospin flavors in order to determine the superconducting order parameter. 

\begin{figure*}
\includegraphics[width=0.9\linewidth]{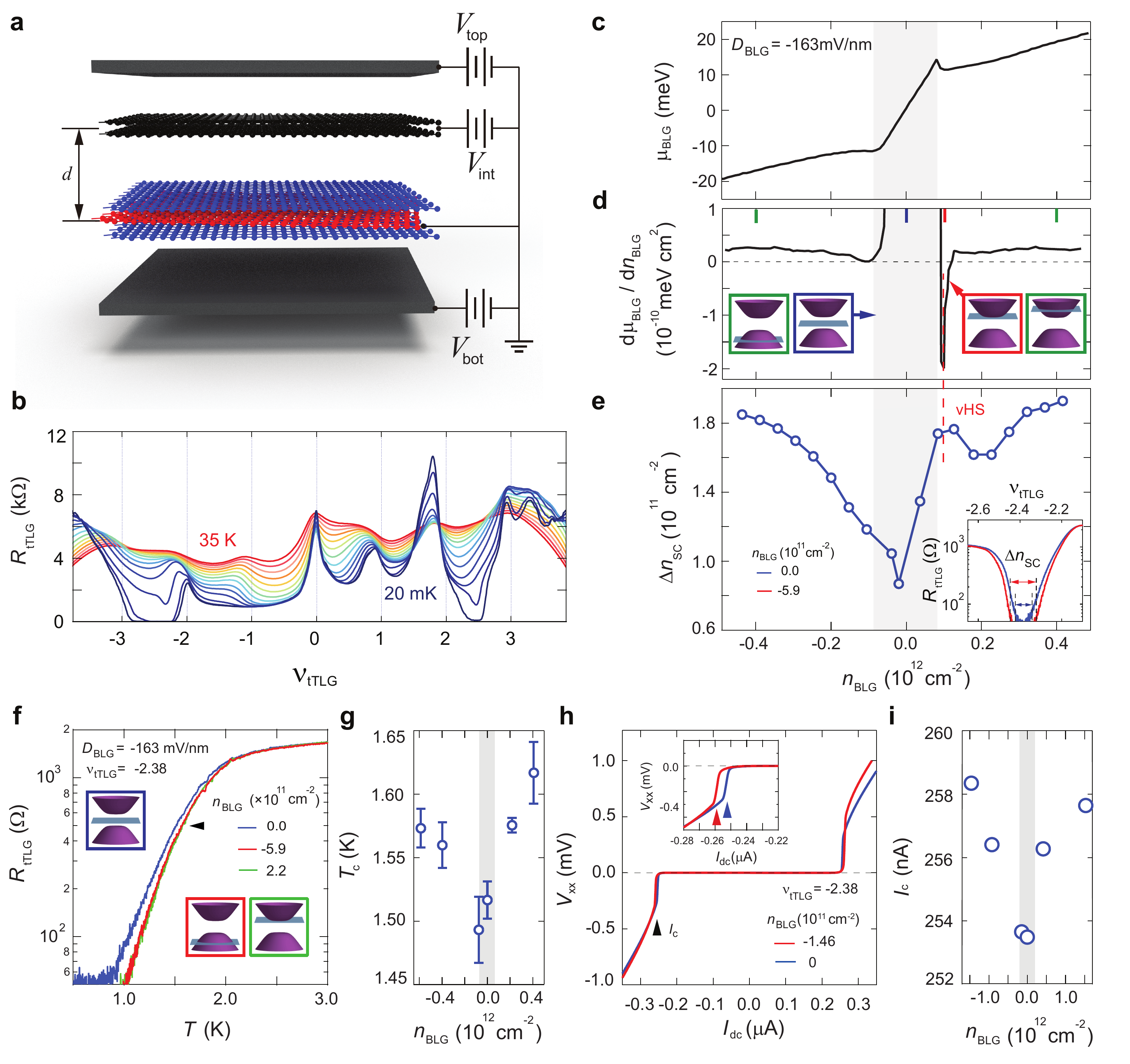}
\caption{\label{figC}{\bf{Magic-angle tTLG in a double-layer structure and Coulomb screening.}}  (a) Schematic of the hybrid double-layer structure consisting of a Bernal BLG and a magic-angle tTLG, separated by a thin insulating barrier with the thickness of $d = 2$ nm. (b) Longitudinal resistance measured from tTLG $R_{tTLG}$ as a function of $\nu_{tTLG}$ at different temperature and  $D_{BLG}$ = 0 mV/nm.  
(c) Chemical potential $\mu_{BLG}$ and (d) inverse compressibility $d \mu_{BLG}/d n_{BLG}$ of BLG as a function of BLG density $n_{BLG}$ at $T = 20$ mK. $\mu_{BLG}$ is extracted from longitudinal resistance $R_{tTLG}$ measured as a function of $V_{int}$ and $V_{top}$. Negative $d\mu_{BLG}/dn_{BLG}$ on the electron edge of the energy gap points toward large density of state associated with an van Hove singularity where screening is enhanced. (e) The density range of the superconducting region $\Delta n_{SC}$ as a function of BLG density $n_{BLG}$ at $T = 20$ mK. Inset: $R_{tTLG}$ as a function of moir\'e filling $\nu_{tTLG}$ measured with different $n_{BLG}$ at $T = 20$ mK and $D_{BLG} = -163$ mV/nm. $\Delta n_{SC}$ is determined by the boundary of the superconducting region, which is defined by the density where $R_{tTLG}$ increases above the noise level. (f) $R_{tTLG}$ as a function of temperature $T$ measured at different $n_{BLG}$ and $D_{BLG} = -163$ mV/nm. The transition temperature $T_c$ is operationally defined as $30 \%$ of extrapolated normal state resistance, marked by the black arrow. (g) $T_c$ as a function of $n_{BLG}$ measured at $D_{BLG} = -163$ mV/nm. At the optimal doping, $D_{BLG}= -163$  mV/nm and $n_{BLG}=0$ correspond to $D_{tTLG}= 95$  mV/nm for tTLG (see Fig.~\ref{nDmapSI} and \ref{inplaneBDmap}). (h) $V_{xx} -I$ curves of the superconducting phase measured at $D_{BLG}$ = 125mV/nm, $T$ = 20 mK and different $n_{BLG}$.   Critical current $I_c$ is defined by the onset in the $V_{xx}-I$ curves,  where the sample transitions from superconducting to normal state behavior.  (i) Critical current $I_c$ as a function of $n_{BLG}$ at $D_{BLG}= 125$  mV/nm, where tTLG experiences $D_{tTLG}= 390$ mV/nm at the optimal doping and $n_{BLG}= 0$. (f-i) Measurements are all performed at the optimal doping of the superconducting phase in tTLG at $\nu_{tTLG}= -2.38$ (see Fig.~\ref{fig:optimal}) ~\cite{SI}. While tTLG is tuned to the optimal doping, in (e), (g) and (i), $D_{tTLG}$ varies by than $\sim 10$ mV/nm over the density range of $n_{BLG}$. The shaded stripes highlight the density range where BLG is insulating.}
\end{figure*}

In this work, we examine two aspects of the superconducting phase in tTLG using a combination of Coulomb screening, thermodynamic and transport measurements.  
By controlling the strength of Coulomb interaction with proximity screening, we show that superconductivity becomes more robust when Coulomb repulsion is suppressed, suggesting that Coulomb repulsion competes against the mechanism underlying Cooper pairing, which is consistent with previous observations in magic-angle tBLG ~\cite{Liu2021DtBLG,Stepanov2019interplay}.  
In addition, we study thermodynamic properties of tTLG using a combination of transport and chemical potential measurements. The thermodynamic energy gap at half moir\'e fillings is shown to be insensitive to a large in-plane magnetic field, which is indicative of a spin-polarized and valley-unpolarized isospin configuration. Furthermore, we examine the phase boundary of the Pomeranchuk-type transition and show that the spin degree of freedom is frozen owing to large stiffness. As a result, valley isospin plays a dominating role in the Pomeranchuk-type phase transition. The large spin stiffness offers further support for the spin-polarized isospin order at half moir\'e filling.
Since the superconducting phase is associated with the fermi surface reconstruction at half-moir\'e filling, identifying the isospin order of the underlying fermi surface provides constraints for the pairing symmetry of superconductivity.

\begin{figure*}
\includegraphics[width=1\linewidth]{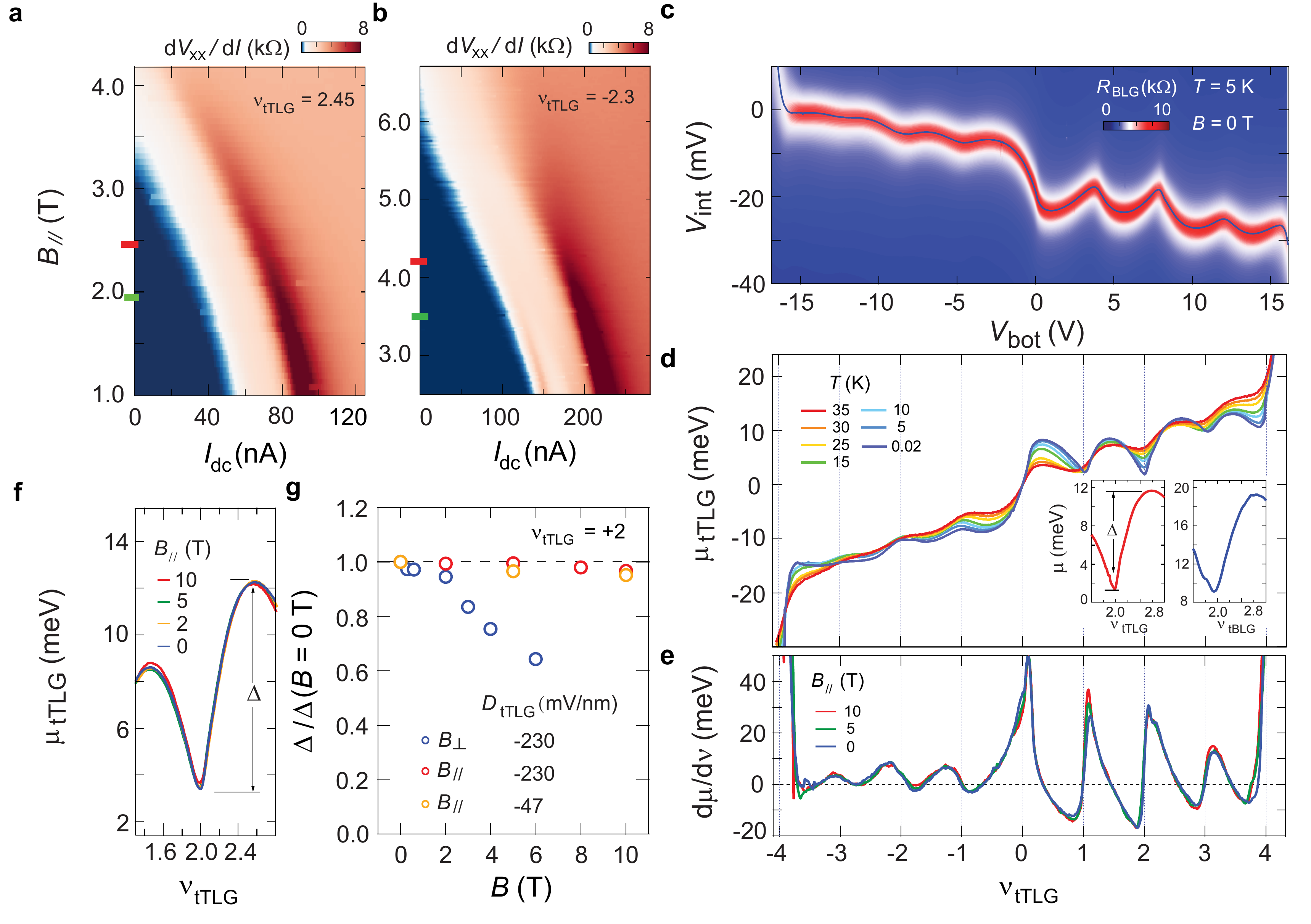}
\caption{\label{figI} {\bf{Isospin orders.}} 
(a-b) Differential resistance $dV_{xx}/dI$ as a function of in-plane magnetic field \Bpara\ measured at (a) $\nu_{tTLG}=2.45$ ($D_{tTLG}$ = -305 mV/nm) and (b) $-2.3$ ($D_{tTLG}$ = 255 mV/nm). The red and green horizontal lines mark the Pauli limit determined from the superconducting transition temperature $T_c$ at $B =$ 0 T, which are defined as 50$\%$ and 10$\%$ of extrapolated normal state resistance, respectively. (c) Longitudinal resistance $R_{BLG}$ measured from Bernal BLG as a function of $V_{int}$ and $V_{bot}$. Bernal BLG remains gapless at $D_{BLG} = 0$ throughout the measurement by keeping $V_{top}$ zero. The blue solid curve tracks the CNP of Bernal BLG. (d-e) Chemical potential $\mu_{tTLG}$ at different temperatures, inverse compressibility $d\mu/d\nu$ at different in-plane magnetic field as a function of moir\'e filling $\nu_{tTLG}$. Inset in (d): comparison of chemical potential near $\nu=+2$ measured from tTLG and tBLG using the same method at $T =$ 20 mK. Results from this tBLG sample are also reported in \cite{Liu2021DtBLG} and~\cite{Saito2021pomeranchuk}. The jump in the chemical potential is $10.3$ meV for tTLG, and $10.2$ meV for tBLG. (f) Chemical potential near $\nu_{tTLG}=+2$ at different in-plane magnetic field measured at $T = 5$ K and $D_{tTLG}$ = -230 mV/nm. (g) Energy gap value, extracted from the jump in \muu, as a function of the in-plane and out-of plane magnetic field for different $D_{tTLG}$ (see Fig.~\ref{inplane+2chemical} for more details.).}
\end{figure*}

The geometry of the double-layer structure is shown in Fig.~1a, where a Bernal bilayer graphene (BLG) is placed in proximity with tTLG, separated by an insulating barrier of hexagonal boron nitride (hBN).
By applying voltage bias on top and bottom graphite gate electrodes, $V_{top}$ and $V_{bot}$, as well as across the insulating hBN barrier, $V_{int}$, we are able to independently control  carrier density in BLG and tTLG, $n_{BLG}$ and $n_{tTLG}$, along with the displacement field $D$. Transport measurement in tTLG reveals a series of correlated states appearing at commensurate fillings, which are evidenced by resistance peaks in longitudinal resistance, resets in Hall density, extra Landau fans and the sawtooth pattern in chemical potential and electronic compressibility ~\cite{Cao2018b,Yankowitz2019SC,Lu2019SC,Liu2021DtBLG}.
At the same time, superconductivity emerges as carrier density is detuned from the CI at $\nu_{tTLG} =2+\delta$ and $-2-\delta$ ~\cite{Hao2021tTLG,Park2021tTLG}. We note that commensurate fillings are determined based on the combination of quantum oscillation and chemical potential measurement (see Fig.~\ref{fig:fan} and Fig.~\ref{figI}). Based on the carrier density at commensurate fillings, we obtain a twist angle of $1.5^{\circ}$. Notably, transport behavior from different parts of the sample suggests that twist angle variation across different parts of the sample is on the order of $\pm 0.01^{\circ}$ (Fig.~\ref{fig:device}). The excellent sample homogeneity, combined with the double-layer structure, allows us to study the moir\'e band of tTLG using Coulomb screening and thermodynamic measurements.


First, we utilize Coulomb screening to investigate the role of Coulomb repulsion in stabilizing the superconducting phase. 
Since Coulomb screening from BLG is determined by its electronic compressibility, its strength can be characterized by mapping the position of the CNP in tTLG, which allows us to extract the chemical potential $\mu_{BLG}$ (Fig.~\ref{figC}c) and inverse compressibility $d\mu_{BLG}/dn_{BLG}$ (Fig.~\ref{figC}d) of BLG   as a function of carrier density in BLG $n_{BLG}$ ~\cite{Lee2014BLG}.
According to Fig.~\ref{figC}c-d, large contrast in screening can be achieved by tuning $n_{BLG}$ in the presence of a $D$-induced energy gap ~\cite{Liu2021DtBLG}. When the fermi surface is inside the energy gap at the CNP, which is highlighted by the gray shaded area in Fig.~\ref{figC}c-e, BLG is highly incompressible. In this scenario, screening is absent and Coulomb repulsion within tTLG is maximized. On the other hand, BLG becomes compressible with increasing density $n_{BLG}$ and Coulomb repulsion in tTLG is suppressed by screening. 
As the strength of screening varies with $n_{BLG}$, the stability of the superconducting phase, which is  characterized by the density range of the superconducting region $\Delta n_{SC}$, changes accordingly. Here we define the boundaries of the superconducting region as the density where $R_{tTLG}$ increases above the noise level (inset of Fig.~\ref{figC}e) ~\cite{Liu2021DtBLG}. 
$\Delta n_{SC}$ is at a minimum when BLG is fully incompressible, whereas it increases with increasing $n_{BLG}$ (Fig.~\ref{figC}e). Notably, $\Delta n_{SC}$ exhibits a local maximum near the electron-edge of the energy gap, which coincides with enhanced screening at the van Hove singularity in BLG   (Fig.~\ref{figC}d).

The effect of controlling Coulomb screening is also observed in the critical temperature $T_c$ and critical current $I_c$ of the superconducting phase at the optimal doping (See Fig.~\ref{fig:optimal}). Both $T_c$ and $I_c$
confirm the same trend as shown in Fig.~\ref{figC}e. 
  The transition temperature $T_c$, operationally defined as $30 \%$ of extrapolated normal state resistance, increases from  $\sim 1.5$ K at $n_{BLG}=0$ to $\sim 1.6$ K when BLG is compressible at large $n_{BLG}$ (Fig.~\ref{figC}f-g). The percentage change in $T_c$ of $\sim 7\%$ is in line with the effect of Coulomb screening observed in magic-angle tBLG ~\cite{Liu2021DtBLG}.
The normal state  resistance at $T\gtrsim 2K$ is shown to be insensitive to changes in the electronic compressibility of BLG (Fig.~\ref{figC}f), demonstrating that changes in impurity scattering does not play a dominating role in the stability of superconductivity  ~\cite{Ponomarenko2011tunable}. 
In addition, the influence of tuning screening onsets near the downturn in the $R-T$ curve, which is characteristic of the emergence of Cooper pairing, suggesting that Coulomb screening is directly influencing the  superconducting energy gap. The $I-V$ characteristics exhibits a similar, albeit less prominent response to variation in Coulomb repulsion:  $I_c$ increases as Coulomb repulsion is suppressed by tuning BLG from insulating to metallic (Fig.~\ref{figC}h-i). 


\begin{figure*}
\includegraphics[width=0.9\linewidth]{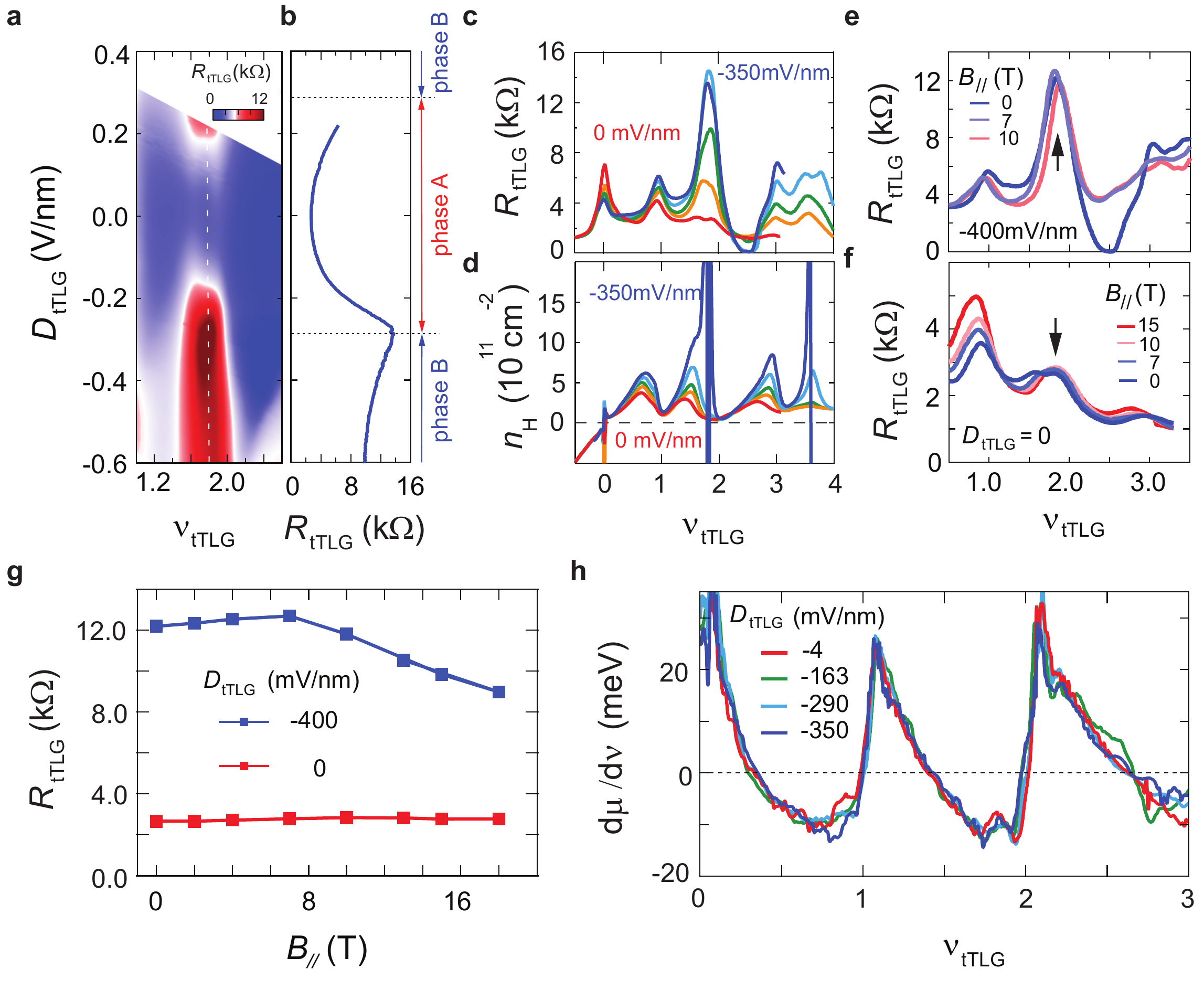}
\caption{\label{figD} {\bf{The effect of displacement field $D_{tTLG}$.}} {(a) Longitudinal resistance in tTLG, $R_{tTLG}$, as a function of moir\'e filling $\nu_{tTLG}$ and displacement field $D_{tTLG}$ at $T =$ 20 mK.  (b) $R_{tTLG}$ as a function of $D_{tTLG}$ measured along the white vertical dashed line in (a). The value of resistance peak reveals two different transport behaviors at half-filling: the resistance peak is suppressed near $D=0$, whereas a prominent peak emerges at large $|D_{tTLG}|> 200$ mV/nm. 
(c) Longitudinal resistance $R_{tTLG}$, (d) Hall density $n_H$ as a function of $\nu_{tTLG}$ at $D_{tTLG}$ = 0, -163, -220, -290 and -350 mV/nm measured at $T=$ 20 mK.  
 (e-f) Longitudinal resistance $R_{tTLG}$ as a function of $\nu_{tTLG}$ at different in-plane magnetic field $B_{\parallel}$ and $T = 300$mK, measured  at (e) $D_{tTLG}=-400$ mV/nm and (f) $D_{tTLG} = 0$. (g) The value of the resistance peak near half-filling marked by the black arrow in (e) and (f) as a function of $B_{\parallel}$ at $T = 300$ mK. (h) Inverse compressibility \dmudnu\ measured from tTLG at different $D_{tTLG}$ and $T=$ 20 mK. $D_{tTLG}$ in (h) is calculated at $\nu_{tTLG}={+2}$.}}
\end{figure*} 


\begin{figure*}
\includegraphics[width=0.9\linewidth]{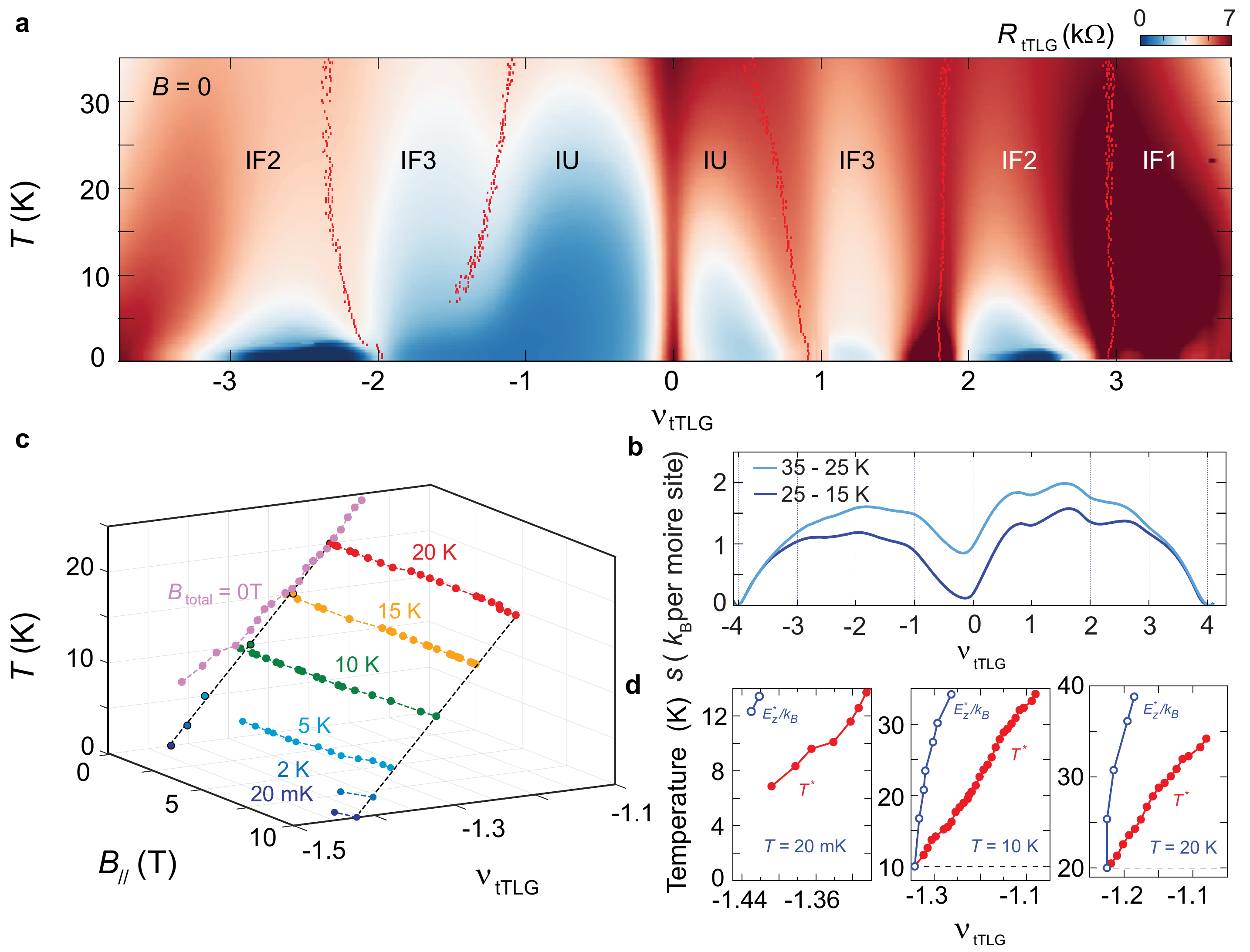}
\caption{\label{figP} {\bf{Isospin Pomeranchuk effect and spin stiffness.}} (a) Longitudinal resistance $R_{tTLG}$ measured from tTLG as a function of $\nu_{tTLG}$ and $T$ at $D_{BLG}$ = 0 mV/nm. Red circles mark the location of local mamxima in $R_{tTLG}$ showing phase boundaries between symmetry-breaking isospin ferromagnets (IF1, IF2, and IF3) and an isospin-unpolarized state (IU) ~\cite{Saito2021pomeranchuk}. (b) Electronic entropy, in units of $k_B$ per moir\'e unit cell, as a function of moir\'e filling $\nu_{tTLG}$ at two different temperatures. Entropy is derived
from the finite difference between $\mu(\nu_{tTLG})$ data measured at 15, 25 and 35 K. Entropy derivative $(\partial S / \partial \nu)_T$ is obtained based on the Maxwell relation $(\partial \mu / \partial T)_\nu = - (\partial S / \partial \nu)_T$, approximating $(\partial \mu / \partial T)_\nu$ from the finite difference of $\mu$ measured at two temperatures. Entropy per moir\'e unit cell is derived by integrating $(\partial S / \partial \nu)_T$ over $\nu$. (c) The position of resistive peaks as a function of $B_{\parallel}$ and T. The black solid circles represent peak position at $B_{\parallel}=10$ T, which are projected to the $B=0$ plane. Dashed curves are polynomial fits to the dots at $B_{\parallel}=10$ T. (d) Phase boundaries for Zeeman-tuned transition ($E_z^{\ast}/k_B$, open blue circles) and temperature tuned transitions ($T^{\ast}$, solid red circles). Boundaries are shown as extracted both from $R_{tTLG}$ peaks. An offset in Zeeman-tuned transitions are included to account for the temperature of the measurement. 
}
\end{figure*}

In all three measurements shown in Fig.~\ref{figC}, the superconducting phase becomes more robust as BLG becomes metallic regardless of charge carrier polarity in BLG. Although varying $n_{BLG}$ introduces small variations in the displacement field experienced by tTLG, $D_{tTLG}$, the robustness of the superconducting phase in all three measurements in Fig.~\ref{figC} varies monotonically with $D_{tTLG}$ (see Fig.~\ref{nDmapSI} and \ref{inplaneBDmap}). Combined with the fact that $D_{tTLG}$ varies monotonically as a function of $n_{BLG}$, according to the capacitance model, the non-monotonic $n_{BLG}$-dependence on Coulomb screening response cannot be accounted for by the influence of varying $D_{tTLG}$. Taken together, we draw the conclusion that the stability of the superconducting phase anti-correlates with the strength of Coulomb repulsion in tTLG, which is consistent with the scenario where Cooper pair formation arises from a mechanism that competes against Coulomb repulsion, such as phonon-mediated coupling  ~\cite{Ochi2018phonon,Lian2019phonon,Wu2018phonon, Liu2021DtBLG}.

Notably, the superconducting phase near $\nu_{tTLG}=\pm 2$ remains robust against a large in-plane magnetic field that exceeds the Pauli limit (Fig.~2a-b). The Pauli paramagnetic limit is defined as $B_{\parallel}^{Pauli}$ = 1.86 [T/K] $\times$ $T_{c}$ (take g = 2), where $T_{c}$ is the critical temperature at $B = $ 0 T  ~\cite{Cao2021Pauli}. This behavior is consistent with the Pauli limit violation in the previous observation ~\cite{Cao2021Pauli}, pointing towards an order parameter that is spin-triplet and potentially unconventional. It is worth pointing out that  in the $B$-field range that exceeds the Pauli limit, the effect of Coulomb screening remains the same as $B=0$, offering further confirmation that the mechanism underlying Cooper pairing  competes against Coulomb repulsion (see Fig.~\ref{screeningbreakPauli}) ~\cite{SI}. However, we note that such pairing mechanism does not offer definitive identification for the superconducting order parameter. To better understand the superconducting phase in tTLG, we will turn our attention to examine thermodynamic properties and the isospin order in the moir\'e flat band.

The double-layer structure allows us to directly extract the chemical potential of tTLG $\mu_{tTLG}$ based on the location of the charge neutrality point in BLG, which reflects the screening of electric field from the bottom gate electrode ~\cite{Lee2014BLG,Park2021flavour}. 
Figure~\ref{figI}c-e plots $\mu_{tTLG}$ and its derivative, inverse compressibility \dmudnu,  as a function of moir\'e filling across the flat energy band. 
The net increase of \muu\ across the moir\'e band provides a direct measurement of the moir\'e bandwidth, which is less than $30$ meV  (Fig.~\ref{figI}d). The measured bandwidth for tTLG is much smaller than tBLG ($40$ meV- $ 60$ meV) ~\cite{Park2021flavour,Saito2021pomeranchuk} and appears consistent with  the calculation for tTLG based on the Bistritzer-Macdonald model ~\cite{Khalaf2019,Bistritzer2011}.
Between each commensurate filling, \muu\ exhibits a decreasing trend that stems from a strong exchange interaction and translates into a negative compressibility of the electron system ~\cite{Eisenstein1994,Lee2014BLG}. A smaller band width  suggests that Coulomb interaction plays a more prominent role in tTLG, which is consistent with the fact that negative compressibility is observed throughout the moir\'e band. 
The series of peaks in \dmudnu\ at each commensurate filling correspond to Fermi surface reconstruction owing to a cascade of isospin-symmetry-breaking transitions (Fig.~\ref{figI}e) ~\cite{Kang2021cascades,Park2021flavour,Zondiner2020cascade,Wong2020cascade}. It is worth pointing out that the behavior of \muu\ and \dmudnu\ of our sample shows an abundance of similarities with that of magic-angle twisted bilayer graphene ~\cite{Park2021flavour,Saito2021pomeranchuk,Rozen2021pomeranchuk}, and the transport response exhibits excellent agreement with previous observation in magic-angle tTLG ~\cite{Park2021tTLG,Cao2021Pauli,Hao2021tTLG}. This indicates that our sample likely has the A-tw-A stacking order ~\cite{CaltecSTMttlg,ColumbiaSTMttlg}.

The ability to probe $\mu_{tTLG}$ allows us to directly examine the isospin order of the fermi surface underlying the superconducting phase at $\nu_{tTLG}=2+\delta$ and $-2-\delta$. This can be achieved by probing the robustness of correlation-driven insulating (CI) state at $\nu_{tTLG}=\pm2$ as a function of in-plane magnetic field \Bpara.
The jump in \muu\ near the half-filling corresponds to the thermodynamic energy gap of the CI $\Delta_{n_s/2}$ (Fig.~\ref{figI}f). 
Notably, $\Delta_{n_s/2}$ is insensitive to in-plane Zeeman coupling up to \Bpara $\sim 10$ T (Fig.~\ref{figI}g), whereas  an out-of-plane $B$ up to $6$ T suppresses  $\Delta_{n_s/2}$ by $\sim 40 \%$. In comparison,   $\Delta_{n_s/2}$ in tBLG exhibits a reduction of $12\%$ to $75\%$ at \Bpara $=9$ T, depending on the measurement methods ~\cite{Yankowitz2019SC,Saito2021pomeranchuk}. Since the spin index couples to both in-plane and out-of-plane $B$ and the valley index is only sensitive to out-of-plane Zeeman coupling,  the distinct responses to different $B$-alignments point towards a spin-polarized, valley-unpolarized isospin configuration for $\nu_{tTLG}=+2$. 

To further confirm the isospin order at half-filled moir\'e band, we examine the transport behavior. 
Figure~\ref{figD}a-d show that varying $D_{tTLG}$ has dramatic effect on both longitudinal resistance and Hall density near $\nu_{tTLG}=2$~\cite{Park2021tTLG,Hao2021tTLG}. The longitudinal resistance $R_{tTLG}$ exhibits a distinguished peak at $\nu_{tTLG} = +2$ and large displacement field $|D_{tTLG}| > 200$, whereas the peak is suppressed at $D_{tTLG}$ = 0. At the same time, a van Hove singularity emerges in the large $|D_{tTLG}|$ regime, evidenced by the diverging Hall density $n_H$ near $\nu_{tTLG}=+2$ (Fig.~\ref{figD}d). 
Notably, the in-plane $B$ dependence of the resistance peak over the entire $D_{tTLG}$ range is consistent with the behavior of the thermodynamic energy gap. In the large $D_{tTLG}$ regime, the resistance peak becomes slightly more resistive up to \Bpara $\sim 7$ T before diminishing slightly  at $10$ T (Fig.~\ref{figD}e), whereas the peak value around $D_{tTLG} = 0$  remains unchanged up to \Bpara $=15$ T (Fig.~~\ref{figD}f). The in-plane $B$-dependence of the resistance peak further confirms the spin polarized and valley unpolarized isospin order at half-filled moir\'e band. If the superconducting phase inherits the isospin order of the underlying fermi surface, a natural order parameter is the spin-triplet, valley singlet ~\cite{Lee2019DBLG,Cornfeld2021}. Such an order parameter is expected to remain robust against an in-plane $B$-field that exceeds the Pauli limit. We note that valley index couples to an in-plane magnetic field through a weak orbital effect ~\cite{Lee2019DBLG}, which could account for the weak $B_{\parallel}$ dependence displayed by $\Delta_{n_s/2}$ ((Fig.~~\ref{figI}g)) and the resistance peak at \Bpara $> 10$ T (Fig.~~\ref{figD}g). At the same time, the orbital effect provides a pair breaking mechanism that destabilizes the superconductivity at large \Bpara. 

It is worth pointing out that varying $D_{tTLG}$ has little influence on the thermodynamic gap (Fig.~2g) and electronic compressibility (Fig.~3h) at $\nu_{tTLG}=+2$. Figure~\ref{figD}h shows that both the location and amplitude of \dmudnu\ peaks remain the same over a wide range of $D_{tTLG}$. The lack of $D_{tTLG}$-dependence in electron compressibility, combined with the emergence of van Hove singularity near $\nu_{tTLG}$ = +2 suggests that isospin-symmetry-breaking transitions are not influenced by the emergence of saddle-points in the Fermi surface ~\cite{Park2021tTLG,Hao2021tTLG}. In addition, the distinct transport response and the robust energy gap in different $D_{tTLG}$ regimes  are indicative of different ground states at half moir\'e filling, tunable with $D_{tTLG}$. For example, a recent theoretical work proposed that the ground state at half-filled moir\'e band in tTLG transitions from an intervalley coherent semimetal around $D_{tTLG} = 0$ to a sublattice polarized insulator at large $D_{tTLG}$, which agrees well with our observations  ~\cite{Christos2021tTLG}.

Lastly, we will characterize a more universal isospin property, spin stiffness, by examining the Pomeranchuk effect in tTLG. Figure~\ref{figP}a plots longitudinal resistance $R_{tTLG}$ in the temperature-moir\'e filling ($T-\nu_{tTLG}$) map. Separated by resistive peaks in $R_{tTLG}$,  different areas in the $T-\nu_{tTLG}$ map  correspond to distinct isospin configurations, which are identified based on the degeneracy of quantum oscillations associated with each commensurate filling (see Fig.~\ref{fig:fan}).
As such, the phase boundary near $\nu_{tTLG}=-1$ marks a temperature driven phase transition from an iso-spin unpolarized (IU) state at low temperature to an iso-spin polarized (IF$_3$) state at high temperature. Such a transition can also be induced at low temperature by applying a large in-plane magnetic field \Bpara, as an extra step in Hall density emerges at $ B_{\parallel} = 10 $T (Fig.~\ref{halldensity}). The duality between the temperature-tuned and Zeeman-tuned transitions between the IU and \IF\ phases points towards a Pomeranchuk-type mechanism~\cite{Saito2021pomeranchuk, Rozen2021pomeranchuk}. In this scenario, the high temperature \IF\ phase is entropically favored  compared to the IU phase owing to fluctuations in local isospin ferromagnetic moments. Notably, the electronic entropy $s$ is extracted based on the temperature dependence of \muu\ and the Maxwell relation  (Fig.~\ref{figP}b). $s$ exhibits a robust minimum around the CNP, rises with both electron or hole doping, reaching a maximum of around $s/k_{B} \sim 1.5$ between $\nu_{tTLG}=1$ and $2$ before dropping back to zero at $\nu_{tTLG} = \pm 4$. The behavior of $s$, are in line with previous observations in tBLG ~\cite{Saito2021pomeranchuk, Rozen2021pomeranchuk}, highlighting that isospin  fluctuations in these two graphene moir\'e systems are similar. 

Notably, the effect of in-plane Zeeman coupling on the Pomeranchuk transition is much weaker in tTLG.
By defining the transition based on the peak position in longitudinal resistance $R_{tTLG}$, we mark the phase boundary between the IU and \IF\ phases in the $T-B_{\parallel}-\nu_{tTLG}$ space (Fig.~\ref{figP}c). 
Figure~\ref{figP}c shows that the IU-\IF\ phase boundary is mostly insensitive to varying in-plane field up to $B_{\parallel} = $10 T, even though it is highly tunable with increasing temperature.
The effect of thermal and in-plane Zeeman energy, $T^{\ast}$ and $E_z^{\ast}/k_B$, is quantitatively characterized in Fig.~\ref{figP}d.  By comparing thermal and Zeeman energies associated with the same amount of shift in the IU-\IF\ boundary, $\Delta \nu^{\ast}$, we estimate the in-plane Zeeman coupling strength in tTLG to be at least four to eight times weaker compared to thermal effect. This is in stark contrast with previous observation in magic-angle tBLG, where temperature and Zeeman coupling are shown to have similar influence on the IU-\IF\ transition ~\cite{Saito2021pomeranchuk}. 
The weak in-plane Zeeman coupling points towards large spin stiffness, since the spin degree of freedom in tTLG is mostly frozen. As a result, the contribution of valley isospin plays a dominating role in the fluctuating isospin moments at high temperature and the associated electronic entropy.

Similar behavior is observed for the isospin transition near $\nu_{tTLG}=+1$ (Fig.~\ref{fig:P1}), which can be directly compared with previous observations in magic-angle tBLG ~\cite{Cao2018b,Yankowitz2019SC,Rozen2021pomeranchuk}.
A Pomeranchuk-type mechanism is demonstrated by the Fermi surface reconstruction, evidenced by the jump in \muu\ and the sharp peak in \dmudnu, which shifts to smaller filling  with increasing temperature.   At the same time, the position of the same isospin transition appears largely insensitive to in-plane Zeeman coupling, confirming that the spin degrees of freedom is frozen owing to large spin stiffness. 
Large spin stiffness  in tTLG increases the energy cost to form spin skyrmions, making valley skyrmions  energetically more favorable. In the scenario where superconductivity originates from  topological textures  ~\cite{Khalaf2020C2T}, pairing between valley skyrmions is expected to play a more dominating role compared to spin skyrmions.
We anticipate our findings will stimulate future investigations into the isospin order in tTLG and its role in the superconducting order parameter.

\section*{Acknowledgments}
We thank Andrea Young, Oskar Vafek and Yahui Zhang for helpful discussions. This work was primarily supported by Brown University. A portion of this work was performed at the National High Magnetic Field Laboratory, which is supported by the National Science Foundation Cooperative Agreement No. DMR-1644779 and the state of Florida. Device fabrication was performed in the Institute for Molecular and Nanoscale Innovation at Brown University. The authors acknowledge the use of equipment funded by the MRI award DMR-1827453. K.W. and T.T. acknowledge support from the EMEXT Element Strategy Initiative to Form Core Research Center, Grant Number JPMXP0112101001 and the CREST(JPMJCR15F3), JST.

\section*{Competing financial interests}
The authors declare no competing financial interests.

\bibliography{Li_ref}%


\newpage

\newpage
\clearpage

\pagebreak
\begin{widetext}
\section{Supplementary Materials}

\begin{center}
\textbf{\large Coulomb screening and thermodynamic measurements in magic-angle twisted trilayer graphene}\\
\vspace{10pt}
Xiaoxue Liu, Naiyuan James Zhang, K. Watanabe, T. Taniguchi, J.I.A. Li$^{\dag}$\\ 
\vspace{10pt}
$^{\dag}$ Corresponding author. Email: jia$\_$li@brown.edu
\end{center}

\renewcommand{\thefigure}{S\arabic{figure}}
\setcounter{figure}{0}
\setcounter{equation}{0}
\newpage


\noindent\textbf{\uppercase\expandafter{\romannumeral1}. Device Fabrication}

The hybrid double-layer structure used in this study is fabricated by using the ``cut-and-stack'' and dry-transfer technique~\cite{Liu2021DtBLG}. We cut a single monolayer graphene flake into three pieces by AFM before stacking. A poly(bisphenol A carbonate) (PC)/polydimethylsiloxane (PDMS) stamp mounted on a glass slide is used to pick up each layer sequentially. From top to bottom, the sequence of the stacking is: graphite as top gate electrode, $30$ nm thick hBN as top dielectric, Bernal bilayer graphene, $2$ nm thick hBN as the insulating barrier, magic-angle tTLG, $50$ nm thick hBN as bottom dielectric, graphite as bottom gate electrode. The entire structure is released onto a Si/SiO$_2$ substrate. The hybrid double-layer stack is shaped into an aligned Hall bar geometry, as shown in Fig.~\ref{fig:device}a. In this geometry, electrical contacts to both Bernal bilayer and tTLG are made independently by the reactive ion etching of CHF$_3$/O$_2$ and deposition of the $Cr/Au$ (2/100 nm) metal edge contacts, which enables the independent electrical measurements in tTLG and Bernal bilayer graphene.

To determine the carrier density associated with accurate integer filling, we use two independent methods in this work: (i) integer filling can be identified by extrapolating quantum oscillations associated with fermi surface reconstruction to $B$ = 0; (ii) track the sawtooth pattern in chemical potential as a function of density. First, we use the quantum oscillations emanating from the charge neutral point and $\nu_{tTLG} = \pm 2$ to identify the location of the charge neutrality point and the carrier density at half moiré filling, n ($\nu_{tTLG} = \pm 2$). Assuming the density at full filling is twice that of half-filling, n ($\nu_{tTLG} = \pm 4$) =2* n ($\nu_{tTLG} = \pm 2$), we calculate the twisted angle $\theta$ based on this equation: $n(\nu_{tTLG} = \pm 4)$ = 8$\theta^{2} / \sqrt{3} a^{2}$, where $a$ = 0.246 nm is the lattice constant of graphene. Secondly, we double check this calculation by comparing with the sawtooth pattern in chemical potential as a function of carrier densities in tTLG, which confirms a twist angle of 1.5 degrees with an error bar of ~0.01 degree.

\begin{figure}[h]
\includegraphics[width=0.95\linewidth]{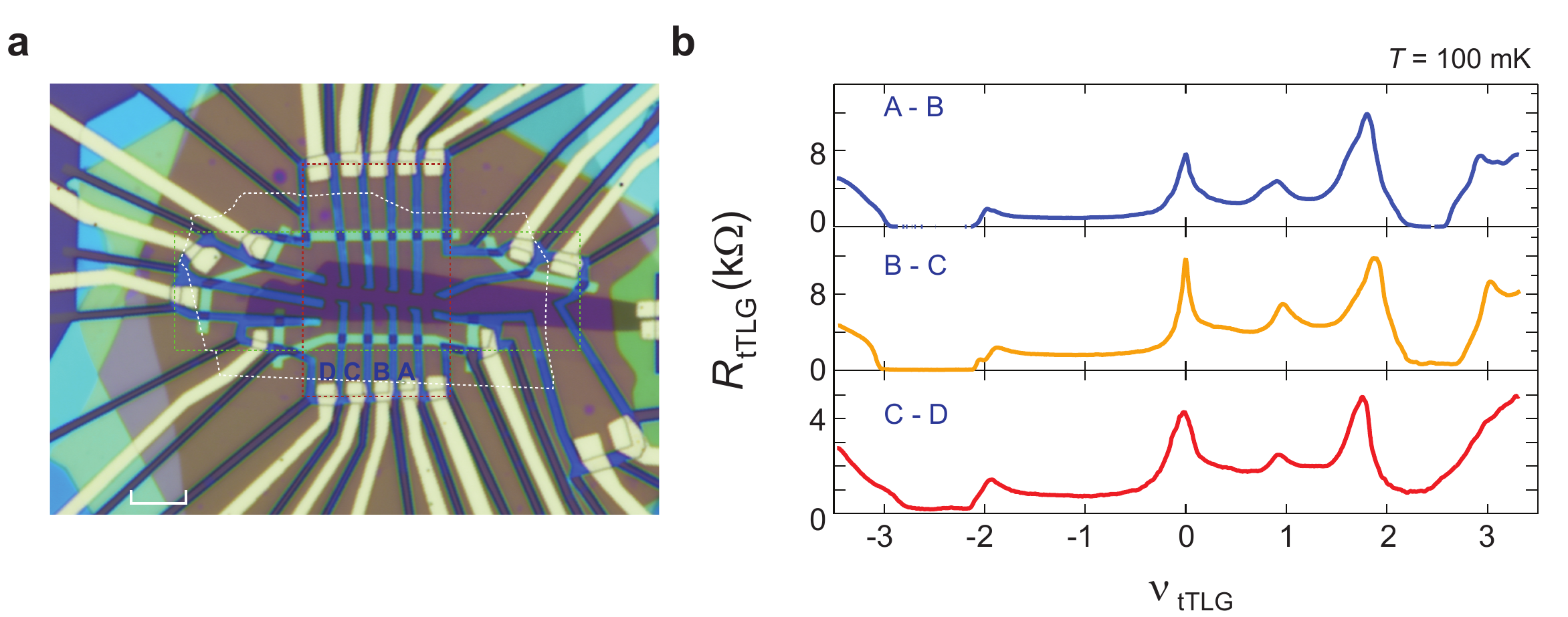}{S}
\caption{\label{fig:device} {\bf{The device characterization.}} (a) Optical image of the hybrid double-layer device. The red, green and white dashed contours highlight tTLG, Bernal bilayer graphene and the middle 2 nm thick hBN layers, respectively. The hall bar channel is fabricated in the bubble-free region. The scale bar is 5 $\mu$m. (b) The four-terminal longitudinal resistance $R_{tTLG}$ obtained from different contact pairs labelled in (a) vs $\nu_{tTLG}$ at $T = $100 mK. The measured $R_{tTLG}$ are almost the same among different contact pairs, showing the high uniformity in this device.
}
\end{figure}

\noindent\textbf{\uppercase\expandafter{\romannumeral2}. Transport measurements}

The device geometry of the hybrid double-layer structure allows independent control of carrier density in Bernal BLG and tTLG, $n_{BLG}$ and $n_{tTLG}$, as well as displacement field $D$. Such control is achieved by applying a DC gate voltage to top graphite electrode $V_{top}$, bottom graphite electrode $V_{bot}$, along with a voltage bias between BLG and tTLG $V_{int}$. $n_{BLG}$, $n_{tTLG}$ and $D$ can be obtained using the following equations: 
\begin{eqnarray}
n_{BLG} &=& (C_{top}V_{top}+C_{int}V_{int})/e+n^0_{BLG}, \label{EqM1}\\
D_{BLG} &=& (C_{top}V_{top}-C_{int}V_{int})/2\epsilon_0, \label{EqM2}\\
n_{tTLG} &=& (C_{bot}V_{bot}+C_{int}V_{int})/e+n^0_{tTLG}, \label{EqM3}\\
D_{tTLG} &=& (-C_{bot}V_{bot}+C_{int}V_{int})/2\epsilon_0, \label{EqM4}
\end{eqnarray} 
\noindent
where $C_{top}$ is the geometric capacitance between top graphite and BLG, $C_{bot}$ the geometric capacitance between bottom graphite and tTLG, and $C_{int}$ the geometric capacitance between BLG and tTLG. $n^0_{BLG}$ and $n^0_{tTLG}$ are intrinsic doping in BLG and tTLG, respectively.

Transport measurement is performed in a BlueFors LD400 dilution refrigerator with a base temperature of $20$ mK. Temperature is measured using a resistance thermometer located on the sample probe. We utilize an external multi-stages low-pass filter that is installed on the mixing chamber of the dilution unit ~\cite{Liu2021DtBLG}.  The filter contains two filter banks, one with RC circuits and one with LC circuits. The radio frequency low-pass filter bank (RF) contains three reflective 7-pole Pi filter stages, which are individually shielded and attenuate above $80$ MHz; whereas the low frequency low-pass filter bank (RC) contains one reflective 7-pole Pi and two dissipative RC filter stages, which are individually shielded and attenuate from $50$ kHz. The filter is commercially available from QDevil. 

We performed measurements in different cryostats with the external magnetic field intended to align with the sample plane. Unless otherwise specified, the tilt angle $\theta$ in this paper indicates the angle between the device plane and the direction of the applied magnetic field. We use Hall resistance as a function of carrier density near the charge neutrality point to determine the tilt angle. Under a certain external magnetic field $B_{total}$, according to $R_{xy}^{-1} = ne/B_{\perp}$, the out-of-plane field component $B_{\perp}$ can be extracted by calculating the linear slope of the inverse Hall resistance $R_{xy}^{-1}$ vs carrier density $n$. The tilt angle $\theta$ is then determined by $B_{\perp} = B_{total}$sin$\theta$.
The first experiment is performed in the BlueFors LD400 system introduced above, where the tilt angle of the sample holder is fixed, resulting in a small tilt angle between the device plane and the applied field of $\sim 2^{\circ}$. 
For this experiment, the tilt angle is determined to be $\sim 2^{\circ} $ (see FIG.~\ref{defineangle}(a)). 
The second experiment is performed in a He3 system with a rotating sample stage at the national high magnetic field lab (SCM-2). We used a single stage low-pass RC filter, which is mounted on the sample probe,  to suppress the influence of the RF noises. To eliminate the out-of-plane component in the $B$-field, we rotate the sample stage at $B_{total}$ = 15 T while monitoring Hall resistance near the charge neutrality point (see FIG.~\ref{defineangle}(b)). 
The magnitude of $R_{xy}$ near the charge neutrality point is minimized when the tilt angle approach zero.  In this experiment, the tilt angle is fixed at zero within the angular resolution that is defined by the minimum step in the mechanical rotation of the sample stage, $\sim 0.1^{\circ}$.  We note that, the re-entrant superconducting phase reported by Ref.~\cite{Cao2021Pauli} is not observed in our device. The absence of the re-entrant phase could arise from several different factors: (i) owing to the accuracy of the mechanical rotator, there is always a non-zero, albeit small, out-of-plane magnetic field. It is possible that the re-entrant phase is suppressed by the out-of-plane component of the $B$-field; (ii) despite the fact that BLG is insulating during the measurement with large in-plane $B$-field, it still contributes extra Coulomb screening which is capable of influencing the ground state order in tTLG ~\cite{Liu2021DtBLG}. It is conceivable that the re-entrant phase is suppressed by the extra Coulomb screening from BLG. If this is the case, our observation would provide strong indication that this re-entrant phase originates from an all-electron mechanism. However, it is important to point out that transport behavior of graphene moir\'e samples often differ from one another owing to sample details such as twist angle distribution and disorder. As such, the absence of re-entrant phase in our measurement cannot be taken as definitive proof of its pairing mechanism, until it is shown by future experiment that the stability of superconductivity anti-correlates with the strength of Coulomb screening. 


Standard low frequency lock-in techniques with Stanford Research SR830 and SR860 amplifier are used to measure resistance $R_{xx}$ and $R_{xy}$, with an excitation current of $1-5$ nA at a frequency of $17.77-43.33$ Hz.

To extract the chemical potential of tTLG $\mu_{tTLG}$, we ground the Bernal bilayer graphene and apply interlayer bias $V_{int}$ to twisted trilayer graphene layer. Voltage bias of $V_{top}$ and $V_{bot}$ are applied to the top and bottom gate, respectively. In this configuration, $n_{BLG}$ and $n_{tTLG}$ are expressed as a function of voltage bias and chemical potentials:

\begin{eqnarray}
en_{BLG}&=& C_{top}\left(V_{top}-\frac{\mu_{BLG}(n_{BLG})}{e}\right)+C_{int}\left(V_{int}+\frac{\mu_{tTLG}(n_{tTLG})-\mu_{BLG}(n_{BLG})}{e}\right)\\
en_{tTLG}&=&C_{bot}\left(V_{bot}-V_{int}-\frac{\mu_{tTLG}(n_{tTLG})}{e}\right)+C_{int}\left(\frac{\mu_{BLG}(n_{BLG})-\mu_{tTLG}(n_{tTLG})}{e}-V_{int}\right),
\end{eqnarray}
\noindent
where $C_{top}$, $C_{bot}$ and $C_{int}$ are geometric capacitance per unit area for the top, bottom and thin middle hBN dieletric layer, respectively. At the charge neutrality point of the Bernal bilayer graphene, $n_{BLG}$=0 and $\mu_{BLG}$=0. Eq. (5) is simplified to: 

\begin{eqnarray}
 \mu_{tTLG}&=& -eV_{int}-e\frac{C_{top}}{C_{int}}V_{top} \\
n_{tTLG}&=&\frac{C_{bot}V_{bot}}{e}+\frac{C_{bot}+C_{int}}{C_{int}e}C_{top}V_{top}
\end{eqnarray}
\noindent
By keeping $V_{top}= 0$,  the value of $V_{int}$ at the charge neutrality point of bilayer graphene offers a direct measurement for the chemical potential in tTLG $\mu_{tTLG}$.  In addition, we extract density $n_{tTLG}$ by $n_{tTLG}$ = $C_{bot}V_{bot}/e$. When fix $V_{top}$ at a constant and non-zero value, the contribution from $V_{top}$ to the chemical potential $\mu_{tTLG}$ is just a constant shift proportional to $V_{top}$. Note that, by fixing the top gate $V_{top}$ at different constant value, the extracted chemical potential $\mu_{tTLG}$ for a constant density $n_{tTLG}$ corresponds to different displacement field $D_{tTLG}$.


\begin{figure}[h]
\includegraphics[width=0.8\linewidth]{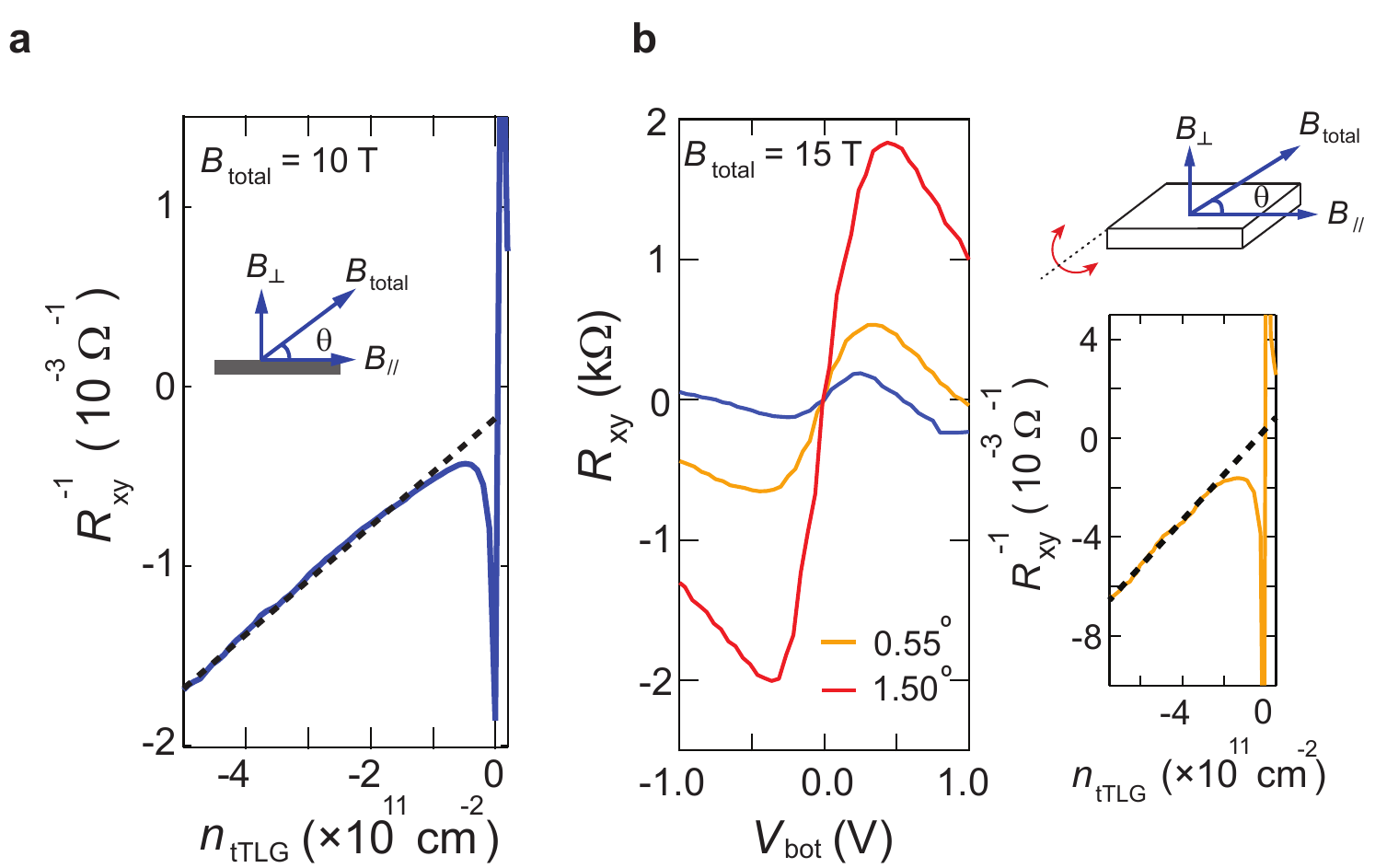}
\caption{\label{defineangle} {\bf{Define the tilt angle for the in-plane magnetic field measurements.}} (a) Inverse Hall resistance as a function of carrier density $n_{tTLG}$ near the charge neutrality point measured at $B_{total} =$ 10 T and $T$ = 20 mK. According to $R_{xy}^{-1} = ne/B_{\perp}$, by calculating the slope from the linear fitting shown as the dashed line, we obtain $B_{\perp}$ = 0.35 T, which corresponds to the tilt angle of $\theta =$ 2$^{\circ}$. (b) Hall resistance $R_{xy}$ versus bottom voltage bias $V_{bot}$ measured with different fixed tilt angles $\theta$ at $B_{total} =$ 15 T and $T$ = 300 mK. The right panel shows the inverse Hall resistance vs $n_{tTLG}$ at $\theta$ = 0.55 $^{\circ}$, and the linear fitting is shown as the black dashed line. The tilt angle is calculated as the same method in (a). The blue trace shows $R_{xy}$ measured at the minimum tilt angle. Given the small Hall resistance, variations in $R_{xy}$ are most likely dominated by mixing from the longitudinal channel.  This minimum tilt angle in our measurement is regarded as the nominal zero tilt angle, where all the fully in-plane magnetic field dependence measurements are performed.}
\end{figure}

\begin{figure}[h]
\includegraphics[width=0.8\linewidth]{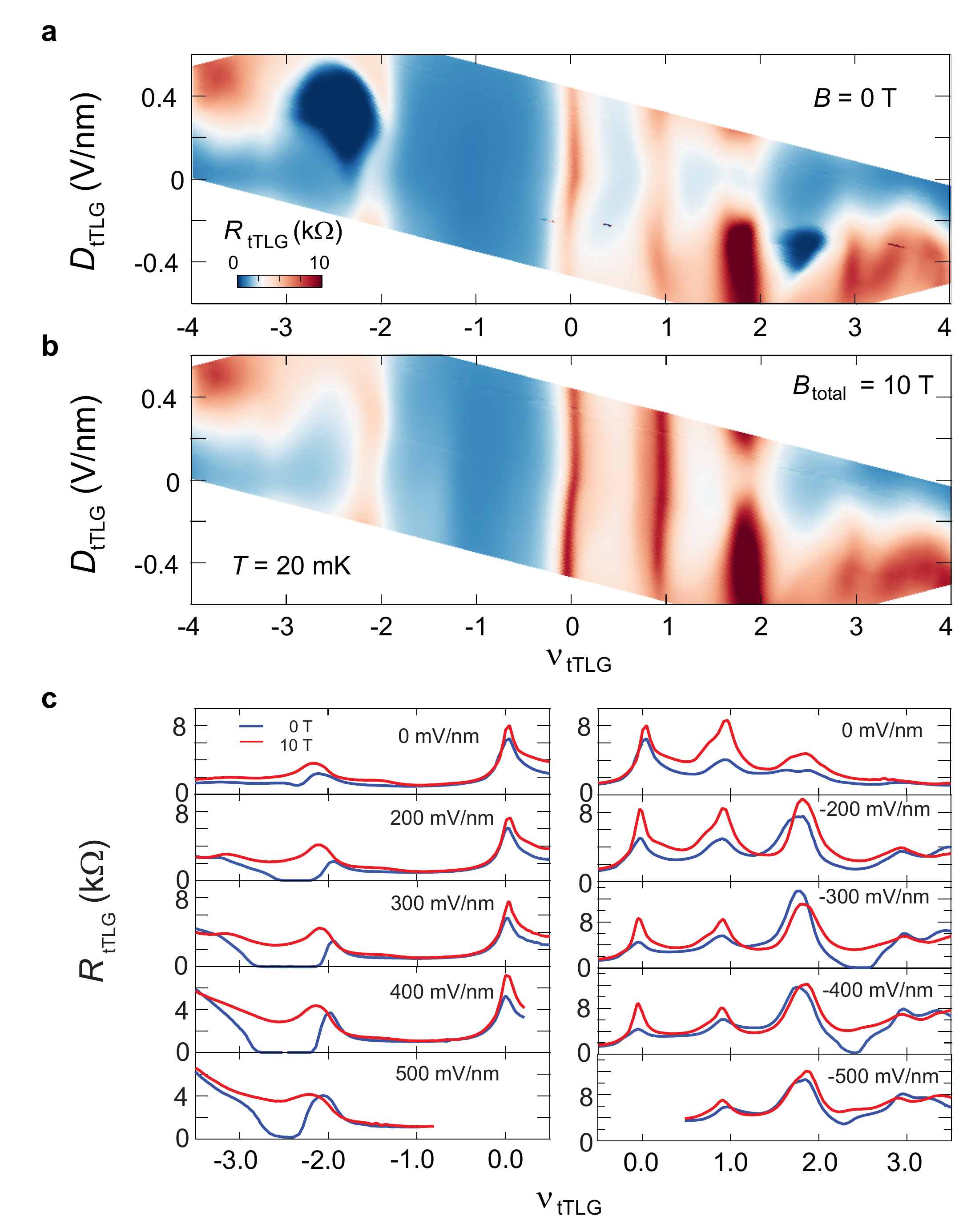}
\caption{\label{nDmapSI} {\bf{The effect of in-plane Zeeman coupling at different \bm{$D$}.}} The longitudinal resistance $R_{tTLG}$ as a function of $\nu_{tTLG}$ and $D_{tTLG}$ at (a) $B = $ 0 T and (b) total magnetic field $B_{total} =$ 10 T measured at $T$ = 20 mK. (c) The linecuts of $R_{tTLG}$ vs $\nu_{tTLG}$ at different total magnetic field $B_{total}$ and $D_{tTLG}$ extracted from (a) and (b). The total magnetic field is oriented at an angle relative to the device plane of $\theta =$ 2$^{\circ}$. }
\end{figure}

\begin{figure}[h]
\includegraphics[width=0.7\linewidth]{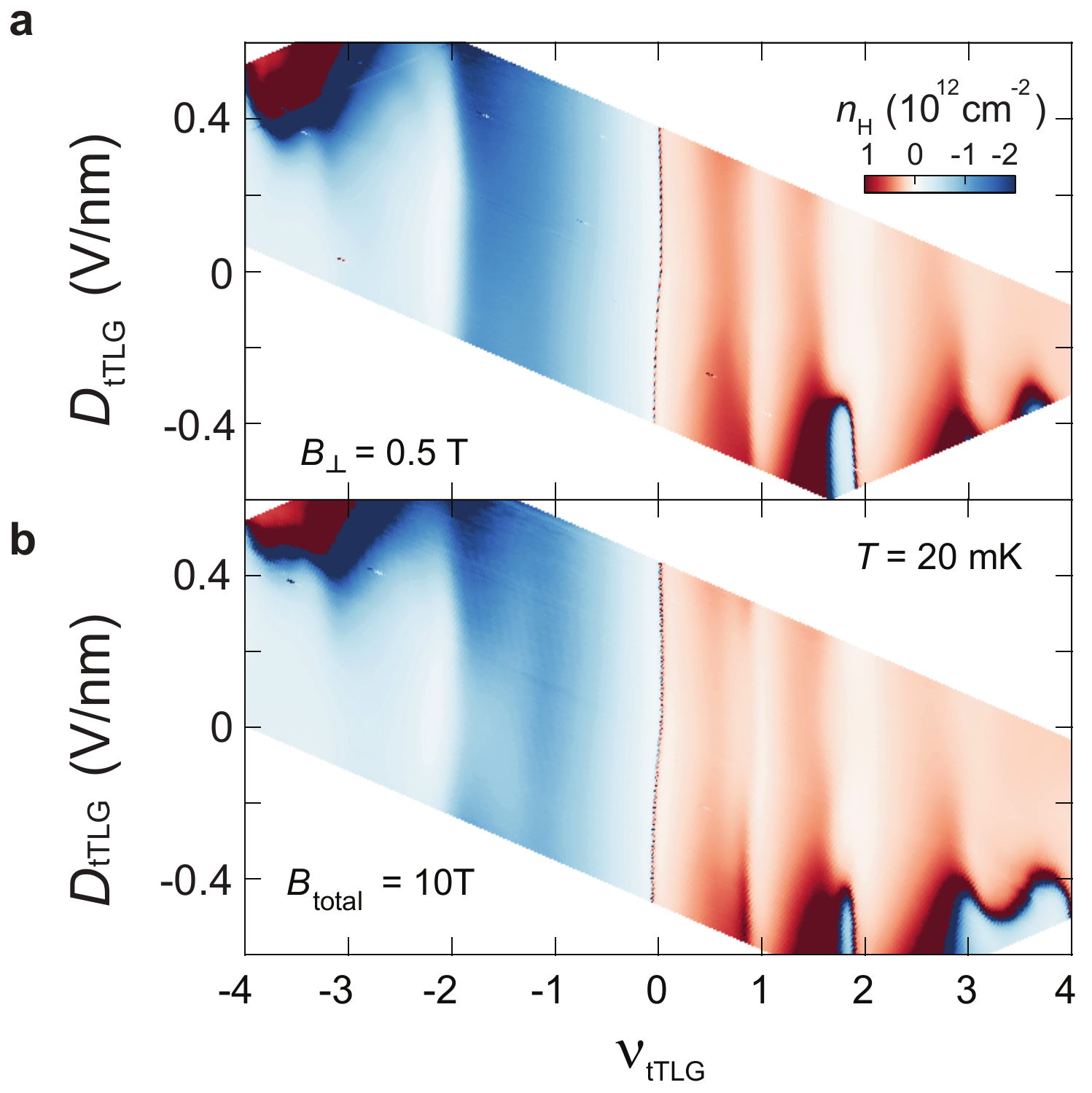}
\caption{\label{halldensity} {\bf{Hall density.}} (a-b) Hall density $n_{H}$ as a funtion of $D_{tTLG}$ and $\nu_{tTLG}$ at $T =$ 20 mK measured at (a) $B_{\perp} = $0.5 T and (b) $B_{total}$ = 10 T oriented at an angle relative to the device plane of $\theta =$ 2$^{\circ}$. Isospin-symmetry-breaking transitions, manifested in Hall density resets, remain largely unchanged in the presence of an in-plane $B$ field.}
\end{figure}

\begin{figure}[h]
\includegraphics[width=0.7\linewidth]{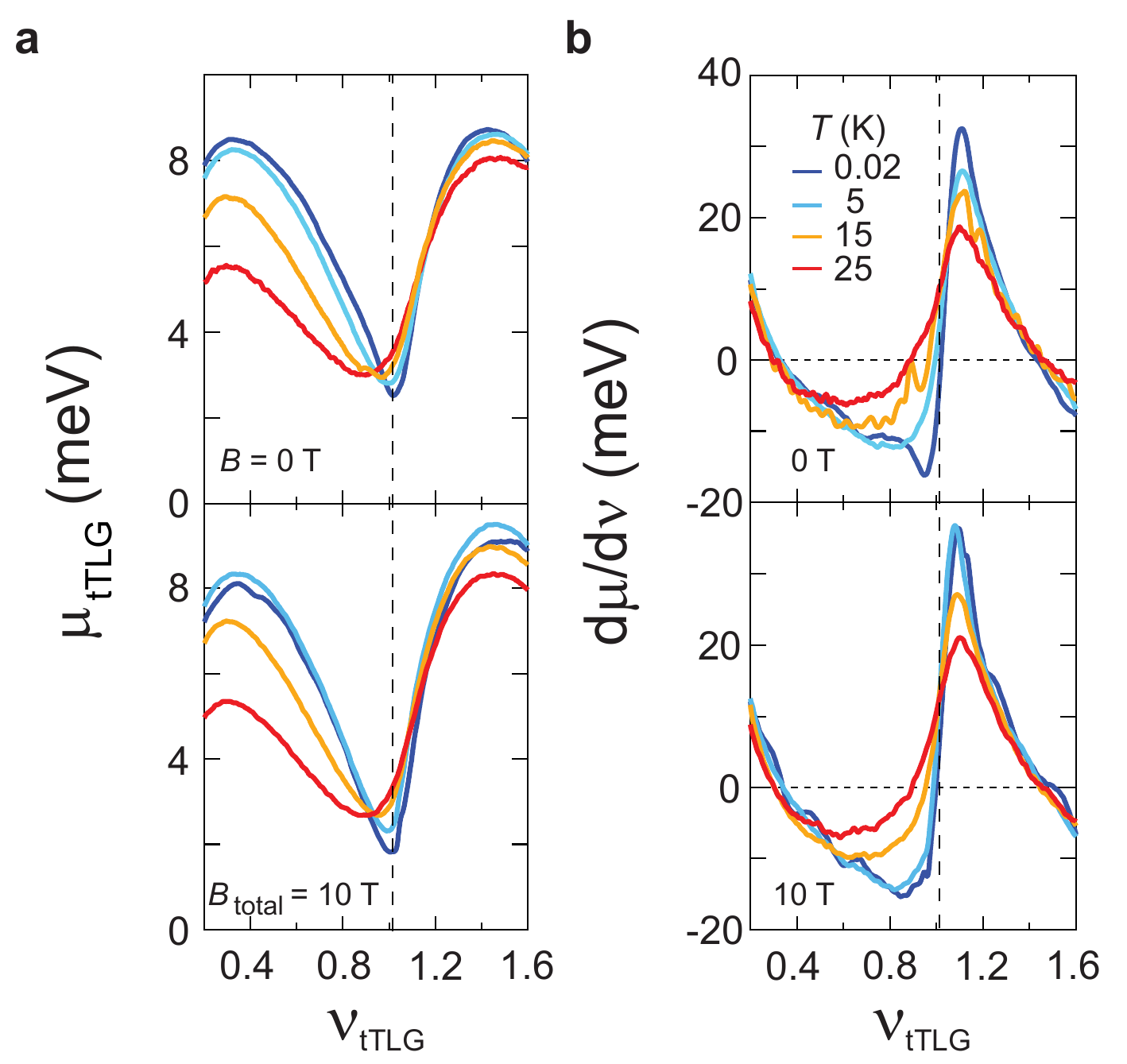}{S}
\caption{\label{fig:P1} {\bf{Pomeranchuk effect near \bm{$\nu=+1$}.}} (a) Chemical potential $\mu_{tTLG}$ and (b) inverse compressibility \dmudnu\ measured near $\nu_{tTLG} = +1$ at $B=0$ (top panel) and $B_{total}=10$ T oriented at an angle relative to the device plane of 2$^{\circ}$ (bottom panel). 
The jump in \muu\ and the sharp peak in \dmudnu\ denote the Fermi surface reconstruction,  which shifts to smaller filling  with increasing temperature.   At the same time, the position of the same isospin transition appears largely insensitive to in-plane Zeeman coupling. This confirms that the spin degrees of freedom is frozen, owing to large spin stiffness, and the Pomeranchuk transition is driven by fluctuations in valley isospin moment. 
}
\end{figure}

\begin{figure}
\includegraphics[width=0.9\linewidth]{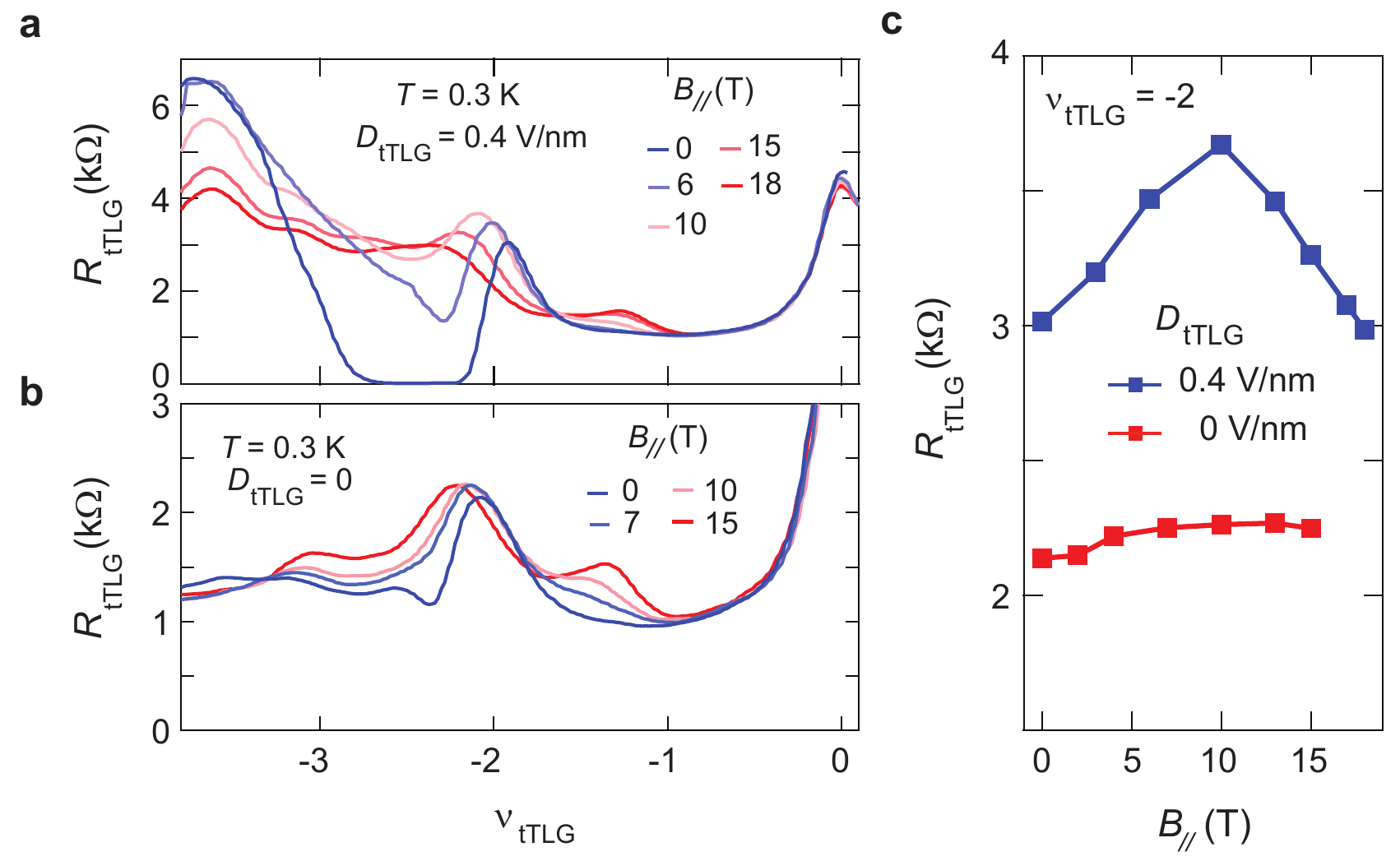}
\caption{\label{m2} {\bf{The in-plane field dependence of $R_{tTLG}$ at $\nu_{tTLG} = -2$.}} $R_{tTLG}$ as a function of $\nu_{tTLG}$ at different in-plane magnetic field $B_{\parallel}$ measured at (a) $D_{tTLG}$ = 400 mV/nm and (b) $D_{tTLG}$= 0 mV/nm, respectively, and $T =$ 300 mK. (c) The value of the resistance peak at $\nu_{tTLG} = -2$  as a function of in-plane magnetic field $B_{\parallel}$ measured at different $D_{tTLG}$ and $T$ = 300mK.}
\end{figure}

\begin{figure}
\includegraphics[width=0.7\linewidth]{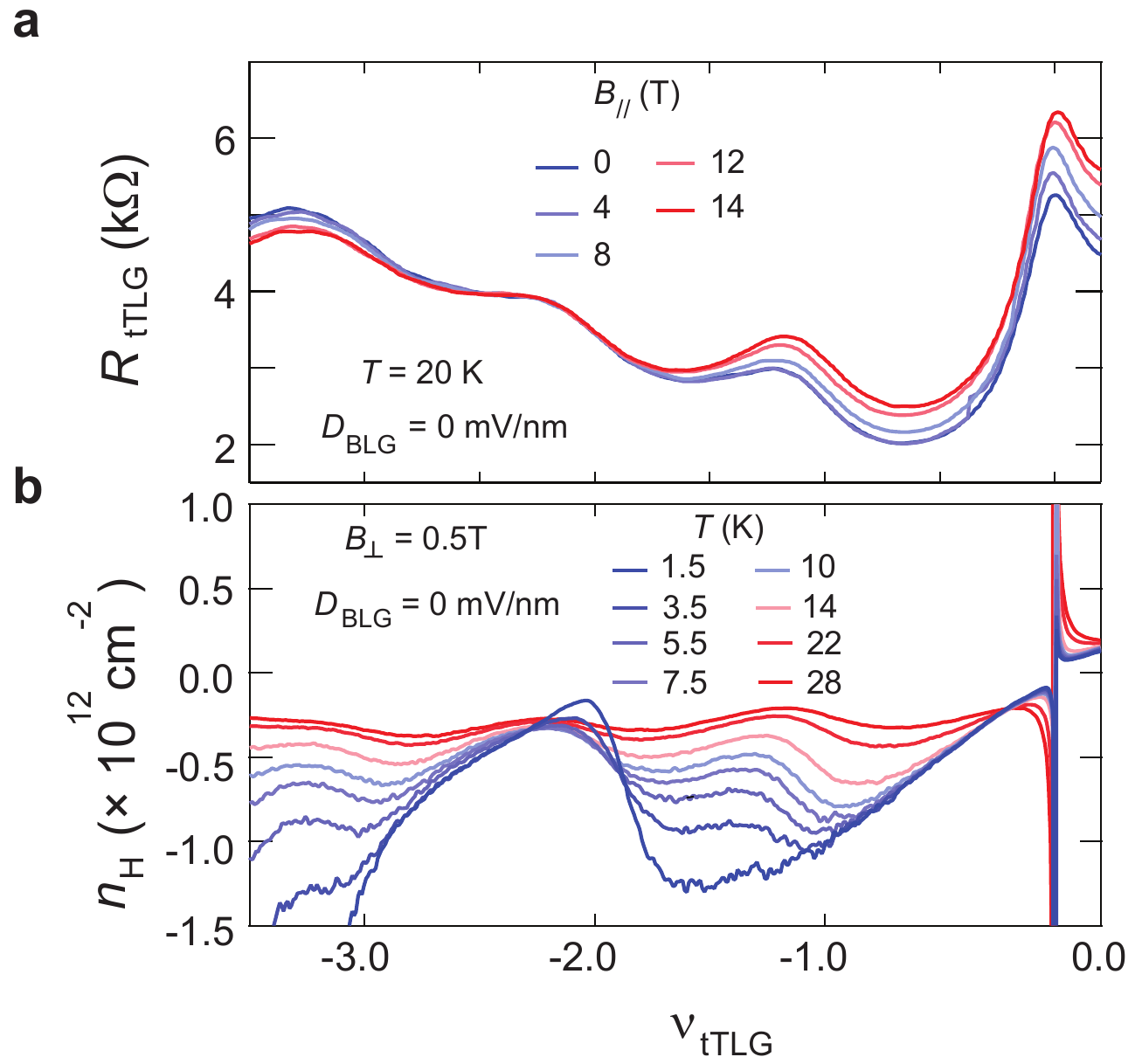}
\caption{\label{Bparapomeranchuk} {\bf{Isospin pomeranchuk effect in tTLG.}} {(a) $R_{tTLG}$ as a function of $\nu_{tTLG}$ at different $B_{\parallel}$ measured at $T =$ 20K and $D_{BLG}$= 0 mV/nm. (b) Hall density $n_{H}$ as a function of $\nu_{tTLG}$ at different temperature measured at $B_{\perp}$ = 0.5 T and $D_{BLG}$ = 0 mV/nm. Near $\nu$ = -1, both the position of the resistance peak in (a) and the emerging kink in Hall density at high $T$ in (b) mark the phase boundary between the symmetry-breaking isospin ferromagnet (IF3) and isospin unpolarized state (IU). Such phase boundary is almost unchanged with tuning $B_{\parallel}$ in (a), while shifts to the charge neutrality point apparently with increasing temperature in (b).}}
\end{figure}

\begin{figure}
\includegraphics[width=0.7\linewidth]{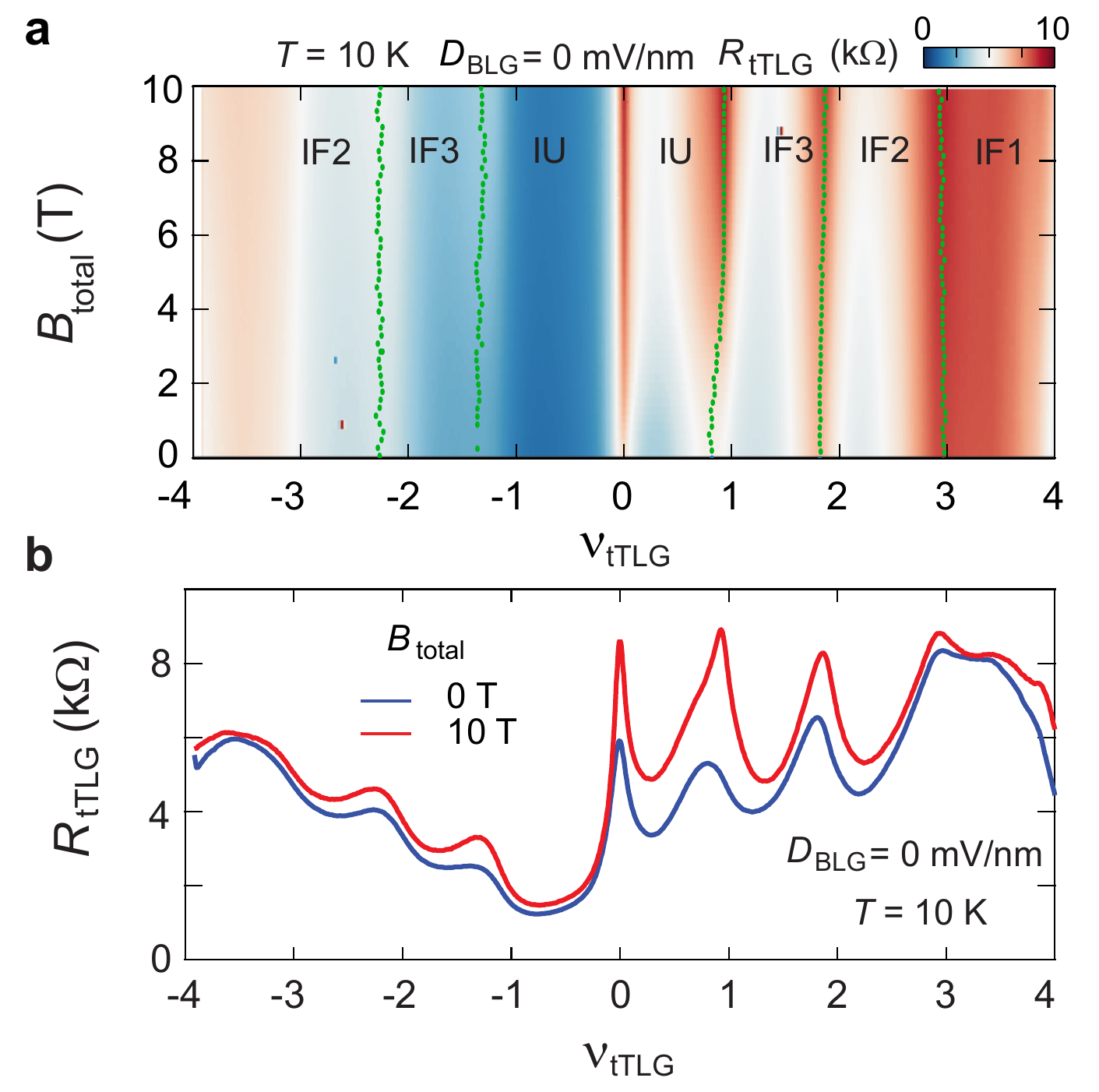}
\caption{\label{BDmapSI}{\bf{The effect of in-plane Zeeman coupling at $10$ K.}} (a) $R_{tTLG}$ as a function of $\nu_{tTLG}$ and $B_{total}$ oriented at an angle relative to the device plane $\theta$  of 2$^{\circ}$ measured at $D_{BLG}$ = 0 mV/nm and $T$ = 10 K. The green dots show the position of $R_{tTLG}$ peak at $\nu=$ $\pm$1, $\pm$2, and +3, which denote the boundaries between an isospin unploarized state IU and the symmetry-breaking isospin ferromagnet IF3, IF2 and IF1. (b) $R_{tTLG}$ versus $\nu_{tTLG}$ at different $B_{total}$  measured at $D_{BLG}$ = 0 mV/nm and $T$ = 10 K extracted from (a). The total magnetic field is oriented at an angle relative to the device plane of $\theta =$ 2$^{\circ}$.}
\end{figure}




\begin{figure}
\includegraphics[width=0.7\linewidth]{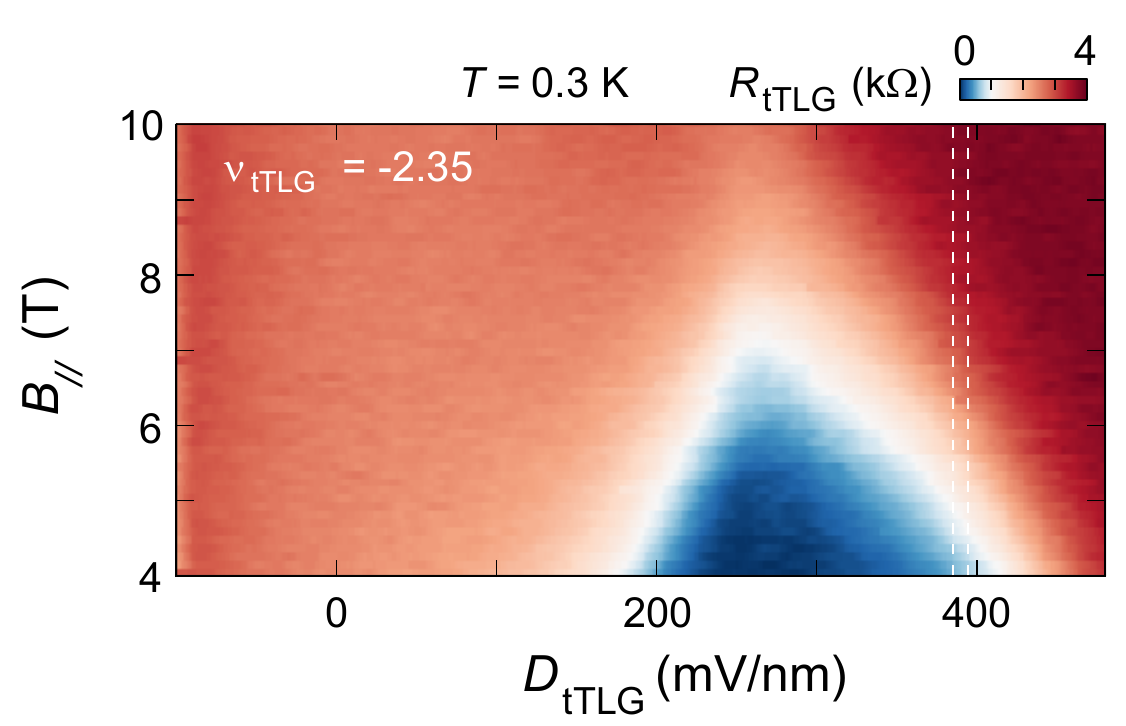}
\caption{\label{inplaneBDmap}{\bf{The $D_{tTLG}$ dependence of superconductivity near $\nu_{tTLG}$ = -2 in the Pauli-limit violation regime.}} $R_{tTLG}$ as a function of $D_{tTLG}$ and $B_{\parallel}$ measured at $\nu_{tTLG}$ = -2.35 and $T = 300$ mK. The white dashed lines mark the range of $D_{tTLG}$ variation induced by changing $n_{BLG}$ in Fig. 1(i).}
\end{figure}

\begin{figure}
\includegraphics[width=0.7\linewidth]{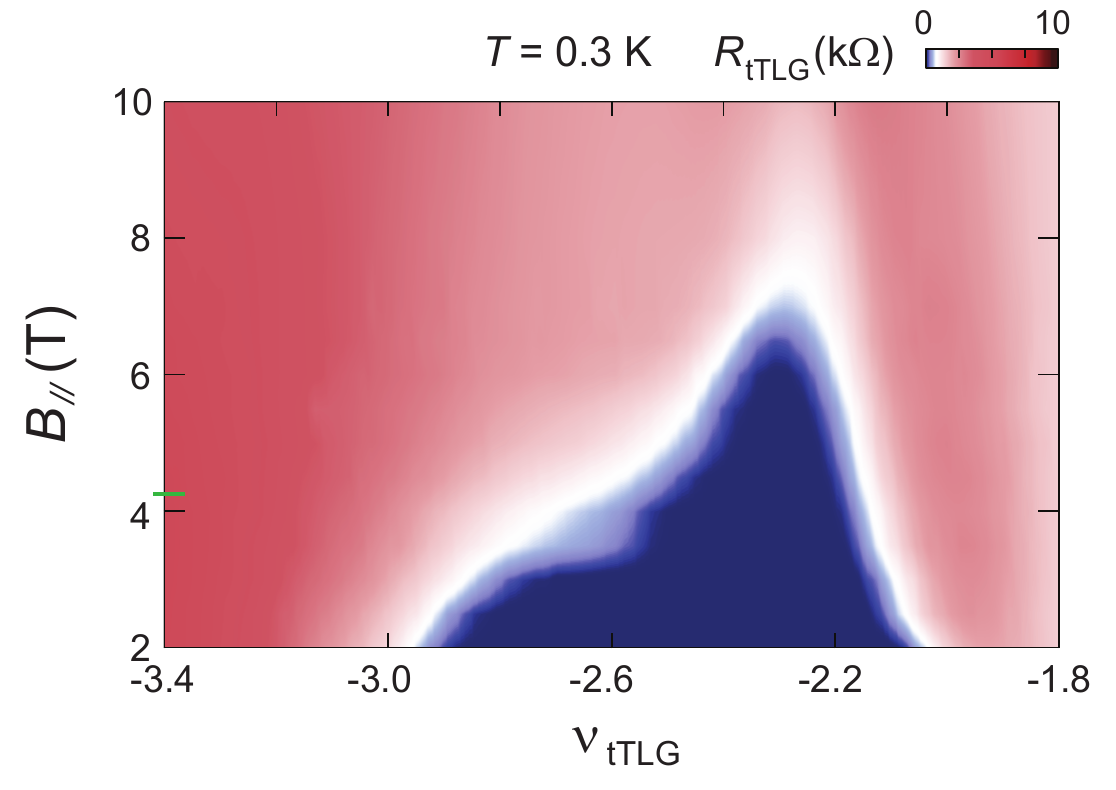}
\caption{\label{0mVTmapbreakPauli}{\bf{The in-plane magnetic field dependence of superconductivity near $\nu_{tTLG}$ =-2.}} $R_{tTLG}$ as a function of $\nu_{tTLG}$ and $B_{\parallel}$ measured at $T =$ 300 mK and $D_{BLG}$ = 0 mV/nm. The green tick marks the Pauli limit at the optimal doping $B_{\parallel}^{Pauli}$, which is defined by $B_{\parallel}^{Pauli}$ = 1.86 (T/K) $\times$ $T_{c}$, where critical temperature $T_{C}$ is defined as 50 $\%$ of the extrapolated normal state resistance at $B$ = 0 T.}
\end{figure}

\begin{figure}
\includegraphics[width=0.7\linewidth]{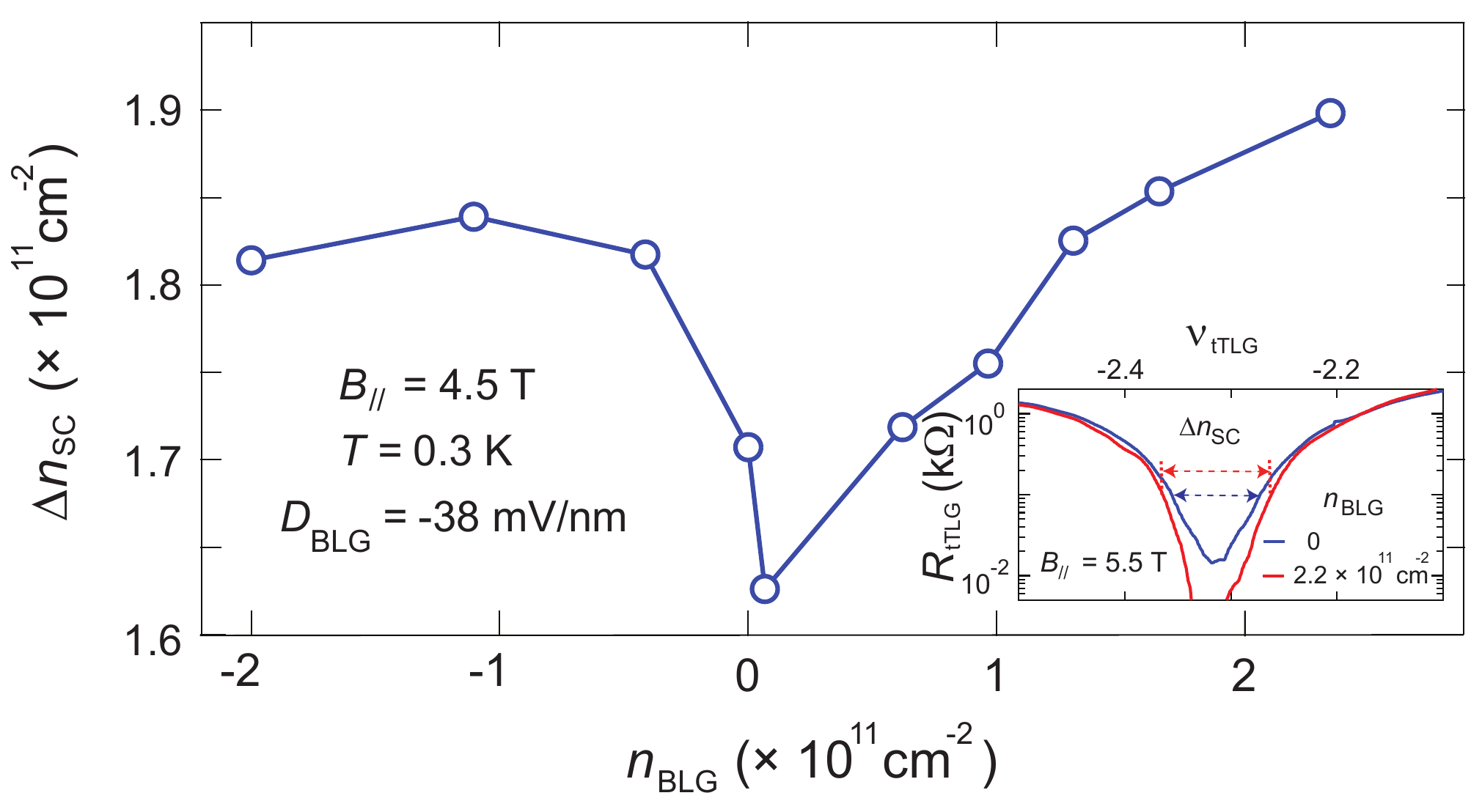}
\caption{\label{screeningbreakPauli}{\bf{The Coulomb screening effect on superconductivity near $\nu_{tTLG}$ = -2 in the Pauli limit violation regime.}} The density range of superconducting region $\Delta n_{SC}$ as a function of Bernal density $n_{BLG}$ at $D_{BLG}$ = -38 mV/nm measured at $B_{\parallel} =$ 4.5 T and $T =$ 300 mK. Due to the Pauli limit at the optimal doping $B_{\parallel}^{Pauli}$ is about 4.2 T, this result is measured at $B_{\parallel} >$  $B_{\parallel}^{Pauli}$, where the pauli limit is violated. The inset shows $R_{tTLG}$ vs $\nu_{tTLG}$ measured at $D_{BLG}$ = -38 mV/nm with different $n_{BLG}$ at $B_{\parallel}$ = 5.5 T and $T = $300 mK. $\Delta n_{SC}$ is determined by the boundary of the superconducting region, which is practically defined by the density where $R_{tTLG}< $100 $\Omega$. At the optimal doping, $D_{BLG} = -$  38 mV/nm and $n_{BLG} = 0$ correspond to $D_{tTLG}$= 205 mV/nm for tTLG. Similar to the results in Fig. 1(e), $\Delta n_{SC}$ in the Pauli limit violation regime is also minimum when Bernal bilayer is fully insulating ($n_{BLG} =$ 0). }
\end{figure}

\begin{figure}
\includegraphics[width=0.7\linewidth]{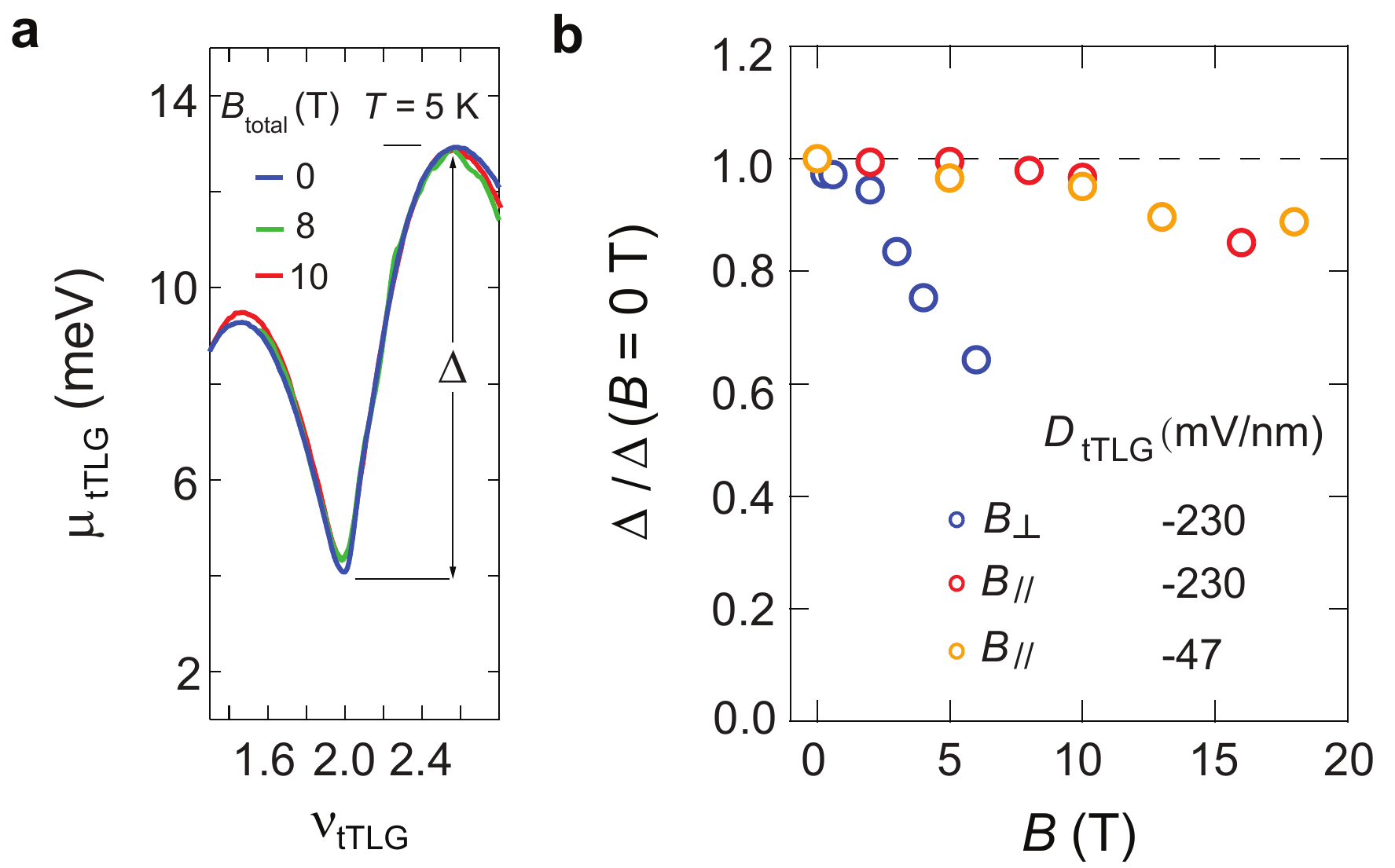}
\caption{\label{inplane+2chemical}{\bf{The magnetic field dependence of the energy gap at $\nu_{tTLG}= +2$ .}} (a) Chemical potential $\mu_{tTLG}$ measured near $\nu_{tTLG}= +2$  with $D_{tTLG}$ = -230 mV/nm at $T$ = 5 K and different magnetic fields. $\mu_{tTLG}$ at $B_{total} = $8 T is measured with $B$ aligned fully in-plane, while $\mu_{tTLG}$ at $B_{total}$ = 10 T is measured with a misalignment of $2^{\circ}$ between $B$ and the sample plane, which correspond to $B_{\parallel} = $10 T and $B_{\perp} = 0.35$ T. The fact that $\mu_{tTLG}$ remains the same indicates that the energy gap is not influenced by a small out-of-plane component of $B$. (b) The same figure as Fig. 2(g). We note that the energy gap at $D_{tTLG} = $-230mV/nm and $B_{\parallel} = $2 T, 5 T, 10 T are measured with a $2^{\circ}$ misalignment between $B$ and the sample plane, giving rise to an out-of-plane component of  $0.35$ T. The rest of the data is measured with $B$ aligned fully in-plane. We note that valley index couples to an in-plane magnetic field through a weak orbital effect ~\cite{Lee2019DBLG}, which could account for the weak $B_{\parallel}$ dependence displayed by $\Delta_{n_s/2}$ and the resistance peak at \Bpara $> 10$ T (Fig.~~\ref{figD}g).}
\end{figure}

\begin{figure}
\includegraphics[width=0.6
\linewidth]{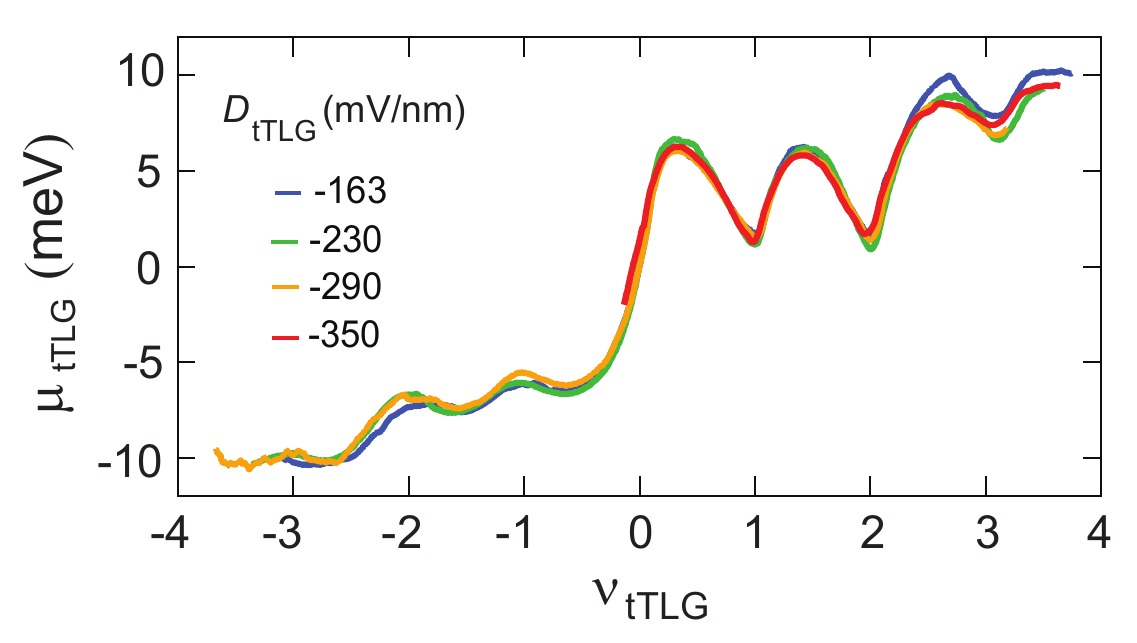}
\caption{\label{bandwidthD}{\bf{Thermodynamic measurement at different $\bm{D}$.}} Chemical potential $\mu_{tTLG}$ as a function of $\nu_{tTLG}$ measured at different $D_{tTLG}$ and $T =$ 20 mK. The labelled displacement field $D_{tTLG}$ denotes $D_{tTLG}$ at $\nu_{tTLG}$ =+2. The chemical potential $\mu_{tTLG}$ at different $D_{tTLG}$ are almost overlapped.}
\end{figure}


\begin{figure}
\includegraphics[width=0.7\linewidth]{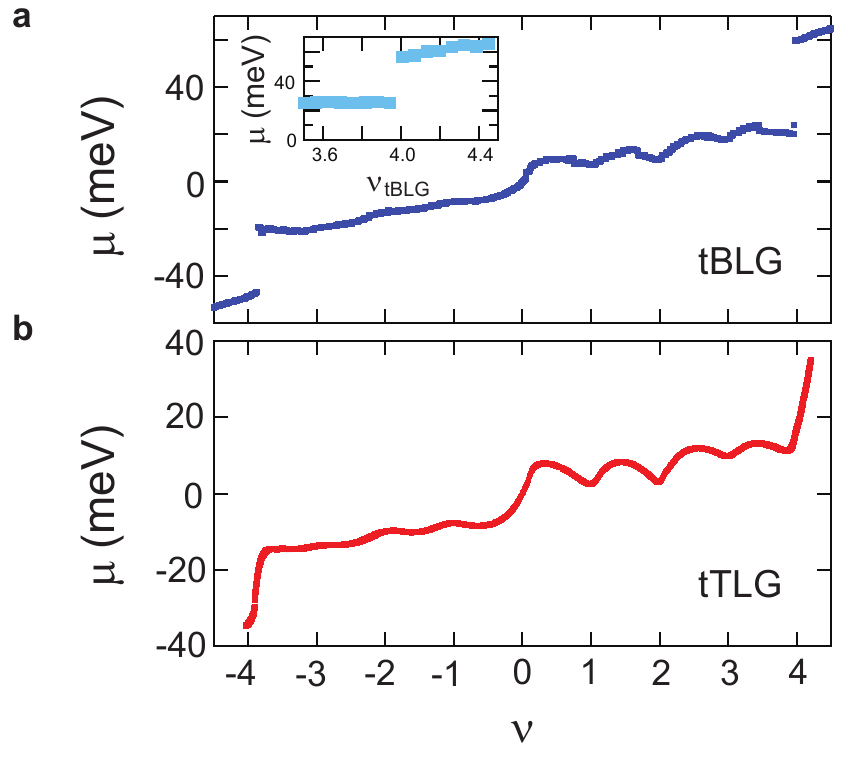}
\caption{\label{fig:chemicalpotential}{\bf{The comparison of chemical potential $\mu$ measured in MATBG and MATTG.}}  {The chemical potential $\mu$ as a function of $\nu$ measured in the magic-angle (a) tBLG (the same device studied in \cite{Liu2021DtBLG}) at $T$ = 20 mK and (b) tTLG studied in this manuscript at $T$ = 5 K, both of which are measured with keeping $V_{top} =$ 0 V. The inset in (a) shows chemical potential $\mu$ vs $\nu$ near $\nu$ = +4 measured at $T$ = 15 K in the same tBLG device. The chemical potential at $\nu$ = +/-4 vary continuously with changing carrier density in tTLG, while it shows the abruptly discontinuous jump in tBLG for both $T$ = 20 mK and 15 K. The different slope of chemical potential in tBLG and tTLG varying with density at $\nu$ = +/-4 reflects the contribution of the monolayer Dirac band in tTLG.}}
\end{figure}

\begin{figure}
\includegraphics[width=0.7\linewidth]{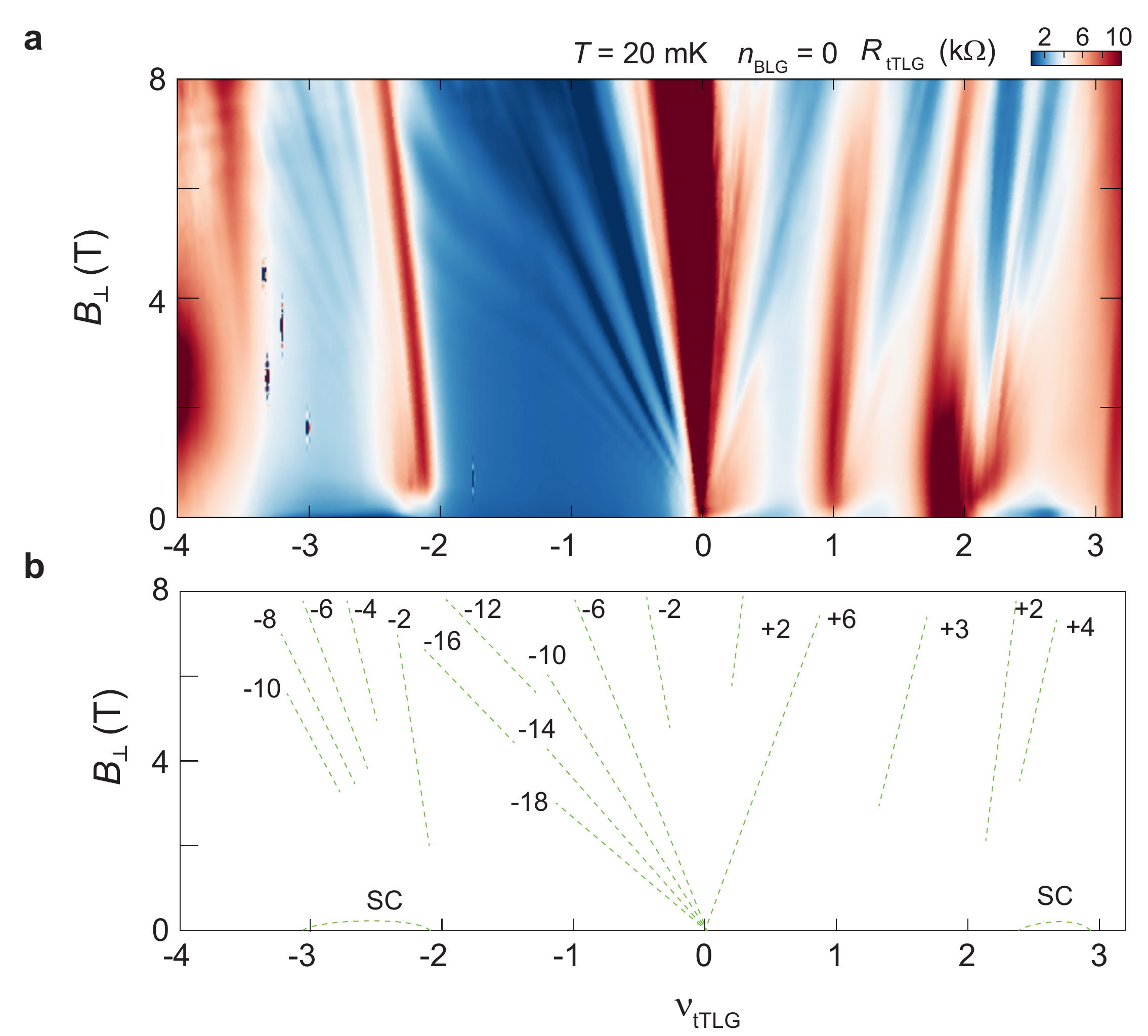}
\caption{\label{fig:fan} {\bf{Quantum oscilation and Landau fans.}} (a) Landau fan diagram of twisted trilayer graphene measured at $D_{BLG} = -50$ mV/nm and $n_{BLG}$ =0 and $T$ =20 mK. (b) Schematic of the landau fan diagram. The number labels the sequence of quantum oscillations emerging from $\nu_{tTLG}$ =0, $+$1 and $\pm$2, separately.}
\end{figure}

\begin{figure}
\includegraphics[width=0.8\linewidth]{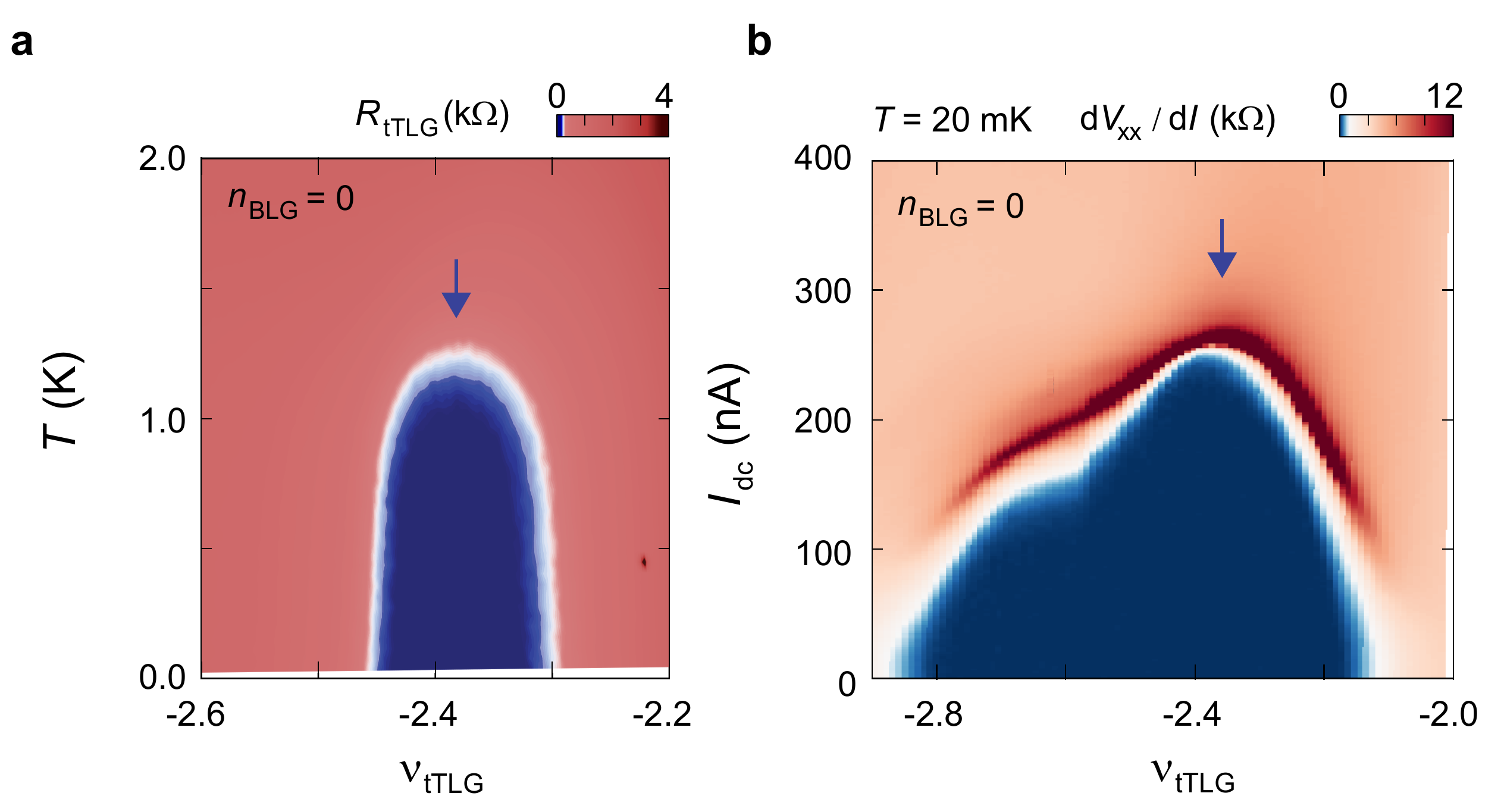}
\caption{\label{fig:optimal}{\bf{Define the optimal doping of superconductivity.}}  {(a) $R_{tTLG}$ vs $\nu_{tTLG}$ and $T$ at $D_{BLG}$ = -163mV/nm with $n_{BLG}$ = 0. (b) $d V_{xx}/d I$ as a function of $\nu_{tTLG}$ and $I_{dc}$ at $D_{BLG}$ = 125mV/nm with $n_{BLG}$ = 0 at T = 20 mK. The arrows mark the optimal doping of the superconducting dome, where the critical temperature and critical current are maximum. The displacement field $D_{tTLG}$ at the optimal doping are (a) $D_{tTLG}$ = 95 mV/nm and (b) $D_{tTLG}$ = 390 mV/nm, respectively.}}
\end{figure}

\begin{figure}
\includegraphics[width=1\linewidth]{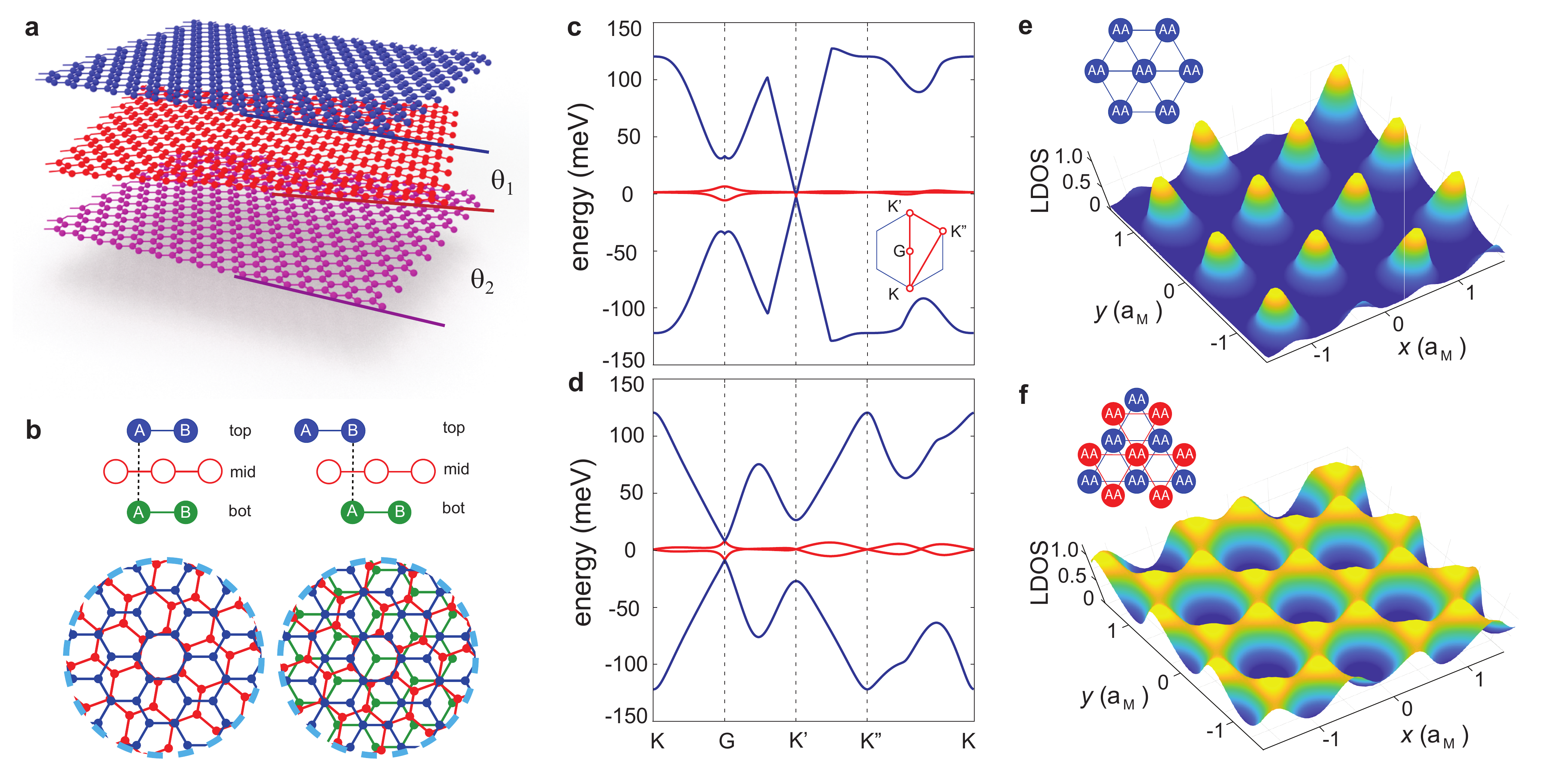}
\caption{\label{fig:AtwBAtwA}{\bf{The comparison of A-tw-A and A-tw-B stacking order.}}  {(a) schematics of a twisted trilayer graphene structure. $\theta_{1}$ and $\theta_{2}$ are twist angles between the top and middle, middle and bottom layers, respectively. (b) top and side views of tTLG with $\theta_{1}$ = $\theta_{2}$ but featuring different stacking orders. Left panel: A-tw-A stacking, the A-sublattice of the top layer is aligned with the A-sublattice of the bottom layer. Right panel: A-tw-B stacking, the B-sublattice of the top layer is aligned with the A-sublattice of the bottom layer. (c-d) calculated band structure for A-tw-A and A-tw-B stacking tTLG near the magic-angle of 1.55 degrees. Both stacking orders feature flat moiré band. (e-f) density distribution of charge carriers in the A-tw-A and A-tw-B stacking tTLG. Charge carriers are concentrated on small islands around the AA sites. These AA sites form a triangular lattice in AA-stacking tTLG, where Wannier function is localized, and Coulomb correlation dominates. In AB-stacking tTLG, AA-sites between top and middle graphene layer are offset from those between the middle and bottom layer (inset in f), thus forming a network with honeycomb lattice. Since electron Wannier function is delocalized in AB-stacking tTLG, we expect Coulomb interaction to play a less prominent role in the ground state order. We thank Yahui Zhang for his input in modeling the band structure and DOS for tTLG with different stacking orders. }}
\end{figure}

\end{widetext}

\end{document}